\documentclass{article}
\usepackage{amssymb}
\usepackage{amsmath}
\usepackage{cases}
\usepackage{stmaryrd}
\usepackage{mathrsfs}
\usepackage{amsfonts}
\usepackage{graphicx}
\usepackage{graphics,color}
\usepackage{setspace}
\usepackage{mathtools}
\mathtoolsset{showonlyrefs=true}

\definecolor{darkgray}{rgb}{0.66, 0.66, 0.66}
 \newtheorem{thm}{Theorem}[section]
 \newtheorem{cor}{Corollary}[section]
 \newtheorem{lem}{Lemma}[section]
 \newtheorem{notn}{Notation}[section]
  \newtheorem{prop}{Proposition}[section]
 \newtheorem{problem}{Problem}[section]
 \newtheorem{defn}{Definition}[section]
 
 \newtheorem{rmk}{Remark}[section]
 \newtheorem{assump}{Assumption}[section]
 \numberwithin{equation}{section}
\newcommand{\filtr}{\mathscr{F}}

\newcommand{\expect}{\mathbb{E}}
\newcommand{\prob}{\mathbb{P}}
\newcommand{\re}{\mathbb{R}}
\newcommand{\ind}[1]{\mathbf{1}_{\left\{#1\right\}}}

\makeatletter
\renewcommand\@biblabel[1]{}
\makeatother

\begin{document}

\begin{center}
\noindent{\huge Facilitation and Internalization Optimal\\ Strategy
in a Multilateral Trading Context}\\
\vspace{30pt}
{\large Qinghua Li}\\
\vspace{20pt} \textit{Humboldt-Universit\"{a}t zu Berlin, Institut
f\"{u}r
Mathematik\\Unter den Linden 6, 10099 Berlin, Germany\\
Email: Ms.QinghuaLi@Gmail.com}\\
\vspace{20pt} {\large \today}
\end{center}
%%%%%%%%%%%%%%%%%%%%%%%%%%%%%%%%%%%%%%%%%
%% \doublespacing
%%%%%%%%%%%%%%%%%%%%%%%%%%%%%%%%%%%%%%%%%
\vspace{20pt} \textbf{Abstract.} This paper studies four trading
algorithms of a professional trader at a multilateral trading
facility, either internalizing or regular, observing a realistic
two-sided limit order book whose dynamics are driven by the order
book events. We shall show that the price switching algorithms
provide lower and upper bounds of the mixed trading algorithms. The
optimal price switching strategy exists and is expressed in terms of
the value function. A parallelizable algorithm to numerically
compute the value function and optimal price switching strategy for
the discretized state process is provided.

\vspace{20pt}

\vspace{10pt}

\section{Introduction}

\subsection{Overview}
Market microstructure is an interdisciplinary field involving
economics, finance, probability and optimization, statistics, and
even psychology, which studies the order-driven price formulation
processes in markets like those of stocks, futures and foreign
exchanges. Due to the complexity of the phenomena, the research
works on market microstructure usually focus on individual aspects
of the problem. Interesting questions studied so far include
econometrics of the order books and of the market maker's inventory
levels, optimal market making, a buyer or seller's optimal order
execution, and limiting behaviors of the queuing
system of limit orders and bid and ask prices.\\
\\The study of market microstructure dates back to at least four decades ago,
and persists up till present time. It is hard to enumerate all the
literature on this field. The books %\cite{OHara1997}
O'Hara (1997) and  %\cite{Hasbrouck2007}
Hasbrouck (2007) provide an overview of quantitative analysis of
market microstructure. One significant development in recent years
is the prevalence of electronic trading platforms as an alternative
to markets where prices are determined via a market maker's auction
and the traders' bidding; the other is the popularity of applying
stochastic control to solving optimal execution and optimal market
making problems. Readers are welcome to Lehalle and Laruelle (2013)
for latest updates in the field
of market microstructure and algorithmic trading. \\
\\Stochastic control provides the theory and methodologies to
find actions that optimize an objective, while the actions can
influence the evolution of some random processes to which the
objective is associated. It naturally facilitates the study of
financial markets where participants, the assembly of whose
activities contribute to the price evolution, seek to maximize
profits and minimize losses. The application of stochastic control
to optimizing activities in an order book traces back to early works
like %\cite{HoStoll1981}
Ho and Stoll (1981).\\
\\There have been many frameworks to study trading and order execution in limit order books.
Among them are the equilibrium models surveyed in %\cite{ParSep}
Parlour and Seppi
(2008), the model with stochastic bid and ask prices
and deterministic order book shape as in %\cite{SchiedSchoneborn2009a},
%\cite{SchiedSchoneborn2009b}, \cite{AlfFruthSchied2010} and
%\cite{AlfSchied2010}
 Alfonsi, Schied and
coauthors (2009, 2010, 2012) and in %\cite{PreShaShr2010}
Predoiu, Shaikhet and Shreve (2011), the model with stochastic mid price and
deterministic or stochastic spread as in %\cite{AveSti2008}
Avellaneda and Stoikov (2008) and %\cite{GuilPham}
Guilbaud and Pham (2013), the Almgren-Chriss model used by many in
the
industry, as in %\cite{Almgren2003}
Almgren (2003), %\cite{AlmChr2000}
Almgren and Chriss (2000), %\cite{BouchardDangLehalle2011}
Bouchard, Dang and Lehalle (2011) and %\cite{GatheralSchied2010}
 Gatheral and Schied (2011), and maximizing
 the utility by choosing an
optimal posting distance that determines the intensity of the
execution process as in  %\cite{GueLehTap2012a}
Gu\'{e}ant, Lehalle and Fernandez-Tapia (2012a, 2012b) %\cite{GueLehTap2012b}
 and %\cite{LauLehPag2011}
Laruelle, Lehalle and Pag\`{e}s (2011).

\subsection{This paper}

An optimal trading scheme is obviously a function of the trading
constraints of the trading agent, translated into its reward
function. To express the adequate trading function, a specific
market model is often needed. Up to now, two main agent types have
been investigated - directional traders and market makers.\\

* Directional traders: such agents already took the decision to buy
or sell and the amount of shares to buy or sell before the trading
phase. Typically institutional investors like pension funds are of
this kind. The control of associated trading schemes is usually the
local trading rate, and more rarely a price (Gu\'{e}ant, Lehalle and
Fernandez-Tapia, 2012a, 2012b). The associated market models include
classical price
diffusion and a market impact component.\\

* Market makers: on the opposite, such agents make decisions in
real time, 100\% based on the state of the order books; they are
simultaneously buyers and sellers, mostly providing liquidity to
other traders. The part of high frequency traders often seen as the
``new middlemen" are of this kind (Baron, Brogaard and Kirilenko,
2012; Jovanovic and Menkveld, 2011; Menkveld, 2013). Their control
is the buying and selling prices, at which they send limit orders
around the best bid and best ask immediate prices. % [ref].
 The associated market models usually embed trading flows abstracted
by a point process, without any market impact component, since the
nature of the market impact of limit orders has not been explored by
now.\\
\\This paper models a third kind of agents: the risk taking
intermediaries. The ``systematic internalizers" defined by the
European regulation are of this kind. Any investment bank having the
capabilities

 (1) to internalize some of its flow against a price
improvement for his external or internal clients;

 (2) to get rid of its potential inventory imbalance, like in a dark pool,
 in a dedicated trading pool at the instantaneous mid price.\\
 The controls will be the number of shares bought and sold up-to-date at every price level in the displayed
 order book and in the dedicated trading pool. Hence the
modeled order book dynamics will have to embed full order book
depth.\\
\\Our study will be presented as follows. Section \ref{sec dynamics}
introduces the event-driven order book dynamics. Sections \ref{sec
trading} and \ref{sec switching} formulate the stochastic control
problems faced by the optimal trader and prove their well-posedness.
Section \ref{sec equiv} compares the best expected profits of a
regular trader and a systemic internalizer, either can use mixed
strategies or price switching strategies. Section \ref{sec sol}
solves the optimal price switching problem by providing a
representation of the optimal strategy, a discrete-time numerical
algorithm and implementation in a Binomial model. Finally, we suggest a way
to calculate a ``fair" internalization premium.\\
\\The contribution of this study is multi-fold.

(1) The \textbf{agent} that conducts the trading activities is a
risk taking intermediary. Such agents make up a significant
proportion of the market participants in terms of the capital
amount, but there has not yet been much research into their optimal
trading strategies.

 (2) One recent development in market microstructure is
the event-driven limit order book \textbf{model}s, by Rama Cont and
co-authors and by Hasbrouck and Saar (2010). Especially when the
trader reacts at a super speed (called ``high frequency trading"),
this kind of models captures the real observations, because the
Central Limit Theorem that proves a diffusion-like stock price no
longer applies. This paper is the first one that derives optimal
trading strategies in a variation of their cutting-edge models.

(3) The optimal \textbf{trading strategy} will balance between the
speed and cost of trading, by active orders in the book and passive
orders at the mid price in the dark pool. There is another kind of
strategies more passive than hidden orders, which is orders queuing
up at the best available prices. Interested readers are invited to
Huang, Lehalle and Rosenbaum (2013) for an empirical analysis and
Lachapelle, Lasry, Lehalle and Lions (2013) for a mean field game
modeling of an agent's optimal queuing.

(4) The \textbf{optimal price switching} problem we shall solve
belongs to the classical type of impulse control and optimal
control, but its state process is non-standard, more complicated
than a textbook SDE driven by Brownian motions and Poisson random
measures.

(5) A \textbf{parallelizable algorithm} is provided for numerically
computing the value function and the optimal price switching
strategy for a discretized state process. The computational
complexity of a stochastic control problem using backward induction
should have been well known on a serial computer, while to the
author's best knowledge this paper is the first one to document the
complexity on a parallel computer.

(6) The results in this paper give insights into trading activities
within the \textbf{Markets in Financial Instruments Directive}
framework, by different types of traders using different types of
trading strategies.

\section{The two-sided order book dynamics}\label{sec dynamics}

As usual in optimal trading (Alfonsi and Schied, 2010, 2012;
Bertsimas and Lo, 1998; Bouchard, Dang and Lehalle, 2011;
Gu\'{e}ant, Lehalle and Fernandez-Tapia, 2012a, 2012b), the market
dynamics are modeled on their own and do not specifically react to
the optimal trader's actions. We will not model any explicit market
impact, following usual frameworks allowing the optimal trader to
post limit orders (Avellaneda and Stoikov, 2008) as opposite to
framework for optimal trading with aggressive orders (Obizhaeva and
Wang, 2013) or at a larger time scale than the orderbook one
(Almgren and Chriss, 2000). Our optimal trader is a ``systematic
internalizer" in the MiFID (Markets in Financial Instruments
Directive) sense: as an intermediary or a dedicated market maker, he
can capture market order flow provided that he pays a premium (i.e.
he improves the price) of the liquidity taker. Our optimal trader
will such implement a trading scheme close to a market making one:
he will capture aggressive flows at the bid and ask, thus earn the
bid-ask spread minus twice the premium he provides. As usual, he
will face a market risk increasing with his inventory (Ho and Stoll,
2008). In our specific case, the optimal trader will operate a Dark
Pool (Cebiroglu, Hautsch and Horst, 2013) or a similar trading
platform, where he will try to unwind its inventory at the
mid-price. High frequency market makers, like Knight Capital Group,
operate such dark pools (``knight link" in this specific
case). %The interaction between the trader and
%the other market participants will be described in section \ref{sec
%trading} and section \ref{sec switching}, depending on the kind of
%trading strategy that the trader uses.

\subsection{Illustration of the order book}
This subsection and the next will introduce the order book dynamics
formed by the aggregate activities of all the market participants,
when the optimal trader does not act.\\
\\In preparation, let us present a few terminologies that appear
frequently in discussions about a limit order book. For every stock
in the market, there are several types of orders, the most commonly
used types being the \textit{limit order} and the \textit{market
order}. A market buy (sell) order only specifies the number of
shares and is executed immediately at the lowest ask (highest bid)
price available in the market. A limit buy (sell) order specifies
the number of shares and the highest (lowest) price at which the
trader is willing to buy (sell). According to the rules of best
price first and FIFO (short for ``first in first out") at the same
price level, limit orders are executed when there are matching sell
(buy) orders at their specified prices. The records of all limit
orders waiting to be executed are maintained. The set of the records
is called a \textit{limit order book}. A limit order book is a
``reservoir'' of limit orders. It records the number of shares, the
price and the time of order arrival or cancelation for every limit
order. Once a limit order is submitted, if it is not executed
immediately, then this order is ``stored'' in the limit order book
until being ``released'' and disappearing from the book for one of
the three reasons -- execution, cancelation, or expiration. The
total of limit orders at each price level is called one
\textit{limit}. The lowest ask price (highest bid price) in the book
is called the \textit{ask price} (\textit{bid price}) for short. The
difference between the ask price and the bid price is called the
\textit{spread}. The distance between two adjacent price levels at
which limit orders can
be submitted is called the \textit{tick size}.\\
\\Fig. 2.1  %\ref{order book}
illustrates a snapshot of a typical limit
order book at some time $t$. The vertical axis represents the
different price levels in the book, where $P^a(t)$ is the current
ask price, $P^b(t)$ is the current bid price and $\delta$ is the
tick size. The horizontal axis represents the volume, in other words
the number of shares, of limit orders at each price level. The sell
side of the book is shown in gray and the buy side in dark. For
example, the volume of limit sell orders at the ask price is denoted
as $Q^a(t)$, which equals the length of the gray horizontal line at
the  price level $P^a(t)$; the volume of limit buy orders at the
price level $P^b(t)-2 \delta$ is denoted as $Q^b_2(t)$, which equals
the length of the dark horizontal line at that level. The spread is
defined as $P^a(t)- P^b(t)$. Without loss of generality, the
tick size is set as $\delta=1$.\\
\\All the limit sell (buy) orders at and higher (lower) than
the best ask (bid) price are displayed to the market participants.
 The volumes of limit orders beyond the best ask and best bid prices are
constants. In the notations illustrated in Fig. 2.1, this means that
$Q^a_1(t)=Q^a_2(t)=\cdots =\Delta^a$ and $Q^b_1(t)=Q^b_2(t)=\cdots
=\Delta^b$, for all $0\leq t\leq T$, where $\Delta^a$ and $\Delta^b$
are two positive constants. The number $Q^a$ of limit sell orders at
the ask price and the number $Q^b$ of limit buy orders at the bid
price are two stochastic processes. When the spread $P^a-P^b$ is
more than one tick, limit sell orders can arrive one tick below the
ask price $P^a$, and limit buy orders can arrive one tick above the
bid price $P^b$. The best ask (bid) price remains constant, until
either all the sell (buy) orders at the current price get depleted
or new limit sell (buy) orders arrive at the price one tick lower
(higher). If the number of all the sell (buy) orders at the best ask
(bid) price reaches zero, then the best ask (bid) price increases
(decreases) by one tick, i.e.
\begin{equation}
P^a(t)=P^a(t-)+1\text{ or }P^b(t)=P^b(t-)-1,
\end{equation}
and the volume at the new ask (bid) price is given by
\begin{equation}\label{volume 2.1}
Q^a(t)=\Delta^a\text{ or }Q^b(t)=\Delta^b.
\end{equation}
If limit sell (buy) orders arrive at time $t$ at one tick below the
ask price $P^a(t-)$ (above the bid price $P^b(t-)$), the ask (bid)
price decreases (increases) by one tick, i.e.
\begin{equation}
P^a(t)=P^a(t-)-1\text{ or }P^b(t)=P^b(t-)+1,
\end{equation}
and each arrival contains $\Delta^a$ ($\Delta^b$) shares, i.e. the
expression (\refeq{volume 2.1}) holds; the number of limit sell
(buy) orders at the old ask price $P^a(t-)$ (the old bid price
$P^b(t-)$) remains $Q^a(t-)$ ($Q^b(t-)$) at time $t$ and resets to
$\Delta^a$ ($\Delta^b$) at time $t+$. We could make
$Q^a_1(t),Q^a_2(t),\cdots$ and $Q^b_1(t), Q^b_2(t),\cdots$ Markov
processes with independent increments. The assumption that they are
constants will significantly reduce the dimensionality of the
control problem while making decisions based on the major driving
forces of the order book dynamics.\\
\\Besides all the displayed orders that form the limit order book in Fig. 2.1,
there is a ``dark pool" mechanism within the spread. Simultaneously,
the trader has the opportunity to place one mid-price pegged order
in the spread: these orders are posted at the price $
\left(P^a(t)+P^b(t)\right)/2 $ at any time $t$. He has to choose if
it is a buy order or sell order, since he cannot simultaneously post
a buy and a sell order at the same price. Other market participants
sending orders of the opposite side and having access to the dark
pool will consume $\Delta^a$ ($\Delta^b$) shares of his order.
%\\Unless specified otherwise, a ``limit order" thereafter refers to a
%``limit order which is displayed to the market participants and is
%not hidden".

\subsection{Mathematical formulation of the order book dynamics}

This subsection will formulate rigorously the dynamics of the limit
order book over a deterministic finite time horizon $[0,T]$.\\
\\Any change to the limit order book, either in the bid and ask
prices, or in the available shares at each price level, is caused by
one of the four types of events -- limit order arrival, limit order
cancelation or expiration, limit order execution, and market order
arrival and immediate execution. The two sources of movements are
the changes in the volumes at the current best prices and the
arrivals within the spread, which in turn result in all the changes
in the prices. The randomness in the order book dynamics is modeled
by the following ingredients.

(1) Probability space $(\Omega,\mathbb{F},\prob)$.

(2) Positive constants $\sigma^a$ and $\sigma^b$.

(3) Independent standard Brownian motions $\sigma^aW^a$ and
$\sigma^bW^b$, representing the evolution of $Q^a$ and $Q^b$ when
there is no price change.

(4) Known measurable functions $\theta^a$, $\theta^b$, $\lambda^a$
and $\lambda^b: \mathbb{N}\rightarrow [0,\infty)$, satisfying
$\theta^a(1)=\theta^b(1)=0$.

(5) Inhomogeneous Poisson processes $N^a$ and $N^b$, with
intensities $\theta^a(P^a(t-)-P^b(t-))$ and
$\theta^b\left(P^a(t-)-P^b(t-)\right)$ at time $t$. When the spread
is larger than one tick, limit sell and buy orders are posted
according to $N^a$ and $N^b$ at a small price improvement - the best
bid plus one tick for a buy order and the best ask minus one tick
for a sell order.

(6) Inhomogeneous Poisson processes $H^a$ and $H^b$, with
intensities $\lambda^a\left(P^a(t-)-P^b(t-)\right)$ and
$\lambda^b\left(P^a(t-)-P^b(t-)\right)$ at time $t$. The trader's
buy and sell orders posted in the dark pool are filled at the mid
price according to the liquidity events $H^a$ and $H^b$.

(7) Conditioning on the spread, the next arrival times of $N^a$,
$N^b$, $H^a$ and $H^b$ are independent of each other and independent
of the future increment of $W^a$ and $W^b$.

(8) The filtration $\filtr=\{\filtr_t\}_{0\leq t\leq T}$, generated
by the processes $W^a$, $W^b$, $N^a$, $N^b$, $H^a$ and $H^b$.\\
\\To prove the well-posedness of
Problem \ref{prob switch} and thus Problem \ref{prob sing}, the
intensities of the order arrival processes within the spread are
assumed uniformly bounded.
\begin{assump}\label{assump 2.1}
The intensity functions $\theta^a$, $\theta^b$, $\lambda^a$ and
$\lambda^b$ of the inhomogeneous Poisson processes $N^a$, $N^b$,
$H^a$ and $H^b$  satisfy
\begin{equation}\label{notn theta star}
\theta^{i*}:=\sup\limits_{p\in \re} \left\{\theta^i(p)\right\} <
\infty \text{ and } \lambda^{i*}:=\sup\limits_{p\in \re}
\left\{\lambda^i(p)\right\}< \infty \text{, }i=a,b.
\end{equation}
\end{assump}
The event-driven limit order book model and the study for an optimal
trading algorithm
based on it are proposed in Section 4.2 by %\cite{Leh2013}
Lehalle (2013). Consistent with existing works, the dynamics indeed
capture the main features of a limit order book. Empirical studies %(\cite{ContStoTal2010}
(Cont, Stoikov and Talreja, 2010, and %\cite{HasSaar2010}
Hasbrouck and Saar, 2010) observe that inhomogeneous Poisson
processes are proper to model the order arrivals and cancelations at
different prices, and that the orders in the neighborhoods closest
to the bid and ask prices being the most influential to the stock
price dynamics. An explanation for the latter observation is that
the limit orders whose execution prices are far away from the bid
and ask prices are more likely to be placed by speculators to profit
from sudden dramatic price changes. Hence, if tracking only the
volumes at the bid and ask prices, it makes a reasonable
approximation to the real limit order books.
%\cite{ContKukSto2010}
The model we use is inspired by Rama Cont and co-authors. Cont,
Kukanov and Stoikov (2014) proposed an \textit{order flow imbalance}
model to describe the stylized features of an order book, where the
number of shares at each price level beyond the best prices is
constant and limit order arrivals and cancelations occur only at the
best bid and ask prices. Further, %\cite{ContLar2010} and \cite{ContLar2011}
 Cont and de Larrard (2011, 2013) have shown that a two-dimensional Brownian motion
 is a reasonable model for the dynamics of the volumes at first limits, when
 the bid-ask spread does not vary too much.\\
\\The number of times over $[0,t]$ that all the orders at the current
ask and bid prices are depleted is
\begin{equation}\label{price 0.3}
L^i(t)= \sum\limits_{0\leq s\leq t}\ind{Q^i(s-) \leq 0}\text{,
}0\leq t\leq T\text{, }i=a,b.
\end{equation}
At every time the volume at the ask (bid) price is depleted, meaning
that $L^i(t)-L^i(t-)=1$, the $\Delta^a$ (respectively $\Delta^b$)
shares at the higher (lower) price level are exposed and the ask
(bid) price increases (decreases) by one tick. At every arrival of
limit sell (buy) orders within the spread, meaning that
$N^i(t)-N^i(t-)=1$, the new limit at the lower (higher) price level
contains $\Delta^a$ (respectively $\Delta^b$) shares  and the ask
(bid) price decreases (increases) by one tick. At any other time,
the volumes move according to the Brownian motions and the prices
remain constants.\\
\\Following the above reasoning, the dynamics of the order book can be
described by the four-dimensional process $(Q^a,Q^b,P^a,P^b)$.  The
\textbf{volumes} $Q^a$ and $Q^b$ move according to
\begin{equation}\label{price 0.2}
\begin{split}
Q^i(t)=  Q^i_0 + \sigma^i W^i(t) + \int_0^t ( \Delta^i - Q^i(t-)
)\,d \,(L^i+N^i)(t)\text{, }0\leq t\leq T\text{, }i=a,b.
\end{split}
\end{equation}
The \textbf{prices} move according to
\begin{equation}\label{price 0.4}
P^a(t)=P^a(0)+L^a(t)-N^a(t)\text{, and
}P^b(t)=P^b(0)-L^b(t)+N^b(t)\text{, }0\leq t\leq T.
\end{equation}
The process $(Q^a,Q^b,P^a,P^b)$ defined in (\refeq{price 0.4}) and (\refeq{price 0.2}) is Markovian.\\
\\Fig. 2.2 and Fig. 2.3 %\ref{plot volume}
plot a simulated path of the two-sided order book dynamics
(\refeq{price 0.4}) and (\refeq{price 0.2}). Fig. 2.2 shows the ask
(top gray line) and bid (bottom dark line) prices. Each time of
price change due to order depletion is assigned a $10\%$ probability
that it is an execution (indicated by circles) and a $90\%$
probability that it is a cancelation. Fig. 2.3
%\ref{plot volume}
shows the volumes at the ask (value of the top gray line) and bid
(absolute value of the bottom dark line) prices respectively in the
positive and negative axis. The parameters are $T=600$, $P^a(0)=20$,
$P^b(0)=15$, $Q^a(0)=Q^b(0)=\Delta^a=\Delta^b=5$,
$\sigma^a=\sigma^b=10$ and
$\theta^a(P^a(t)-P^b(t))=\theta^b(P^a(t)-P^b(t))=0.5(P^a(t)-P^b(t))$.

\subsection{Execution in the dark pool}

This subsection will formulate rigorously the optimal trader's
activities
inside the dark pool.\\
\\The trader's decision on whether to accept an upcoming liquidity
event in the dark pool is indicated by the set of admissible hidden
order strategies
 defined below. Be it an internalizing trader  or a regular trader,
 the admissible set of hidden order strategies is the same.
\begin{defn}\label{adm str 1.2} (hidden order strategy)
The trader's decisions on whether to accept the hidden orders are
indicated by the $\filtr$-adapted, right-continuous,
$\{0,1\}$-valued processes $h^a$ and $h^b: [0,T]\times \Omega
\rightarrow \{0,1\}$. The process $h^a$ equals zero on the set
\begin{equation}\label{adm 1.2.1.a}
\left\{(t,\omega)\in[0,T]\times\Omega | P^b(t)\geq \bar{p}^a
\right\},
\end{equation}
and the process $h^b$ equals zero on the set
\begin{equation}\label{adm 1.2.1.b}
\left\{(t,\omega)\in[0,T]\times\Omega | P^a(t)\leq \underline{p}^b
\right\}.
\end{equation}
At any time $t \in [0,T]$, $h^a(t)$ and $h^b(t)$ cannot both equal
to one. The collection of all such processes $h=(h^a, h^b)$ are
denoted as $\mathscr{H}$. The subset of $\mathscr{H}$ restricted on
$[t_1,t_2]\times\Omega $ is denoted as $\mathscr{H}_{t_1,t_2}$, and
$\mathscr{H}_{t,T}$ is denoted as $\mathscr{H}_t$ for short.
\end{defn}
In Definition \ref{adm str 1.2},  the right-continuity of $h^a$ and
$h^b$ guarantees that, for each scenario $\omega\in \Omega$, the
hidden orders are revised finitely many times over the time horizon
$[0,T]$. %This condition further means that once a hidden order is
%placed, it stays there for at least a short while and is not
%immediately canceled; once a hidden order is canceled, no more
%hidden order is placed in the immediate future.
The value $h^a(t)=1$ ($h^b(t)=1$) means that the trader places a
hidden limit buy (sell) order of $\Delta^a$ ($\Delta^b$) shares with
the execution the price $\left(P^a(t)+P^b(t)\right)/2$ (respectively
$ \left(P^a(t)+P^b(t)\right)/2 $); the value $h^a(t)=0$ ($h^b(t)=0$)
means that he
does not place the hidden order.  %means that once a hidden order is placed, it stays there
%for at least a short while and is not immediately canceled; once a
%hidden order is canceled, no more hidden order is placed in the
%immediate future.
The processes $h^a$ and $h^b$ being zero on the sets in (\refeq{adm
1.2.1.a}) and (\refeq{adm 1.2.1.b}) requires that the trader's
hidden orders would only buy below the price $\bar{p}^a$ and sell above the price $\underline{p}^b$.\\
\\Suppose an liquidity sell (buy) event occurs at time $t$, meaning that
$H^a(t)-H^a(t-)=1$ (respectively $H^a(t)-H^a(t-)=1$). If the trader
placed a hidden limit buy (sell) order right before time $t$ with
the execution price $ \left(P^a(t)+P^b(t)\right)/2 $ (respectively $
\left(P^a(t)+P^b(t)\right)/2 $), then he successfully buys
$\Delta^a$ shares (sells $\Delta^b$ shares)  at time $t$ and pays
(receives) a cash amount of $\Delta^a \left(P^a(t)+P^b(t)\right)/2 $
(respectively $\Delta^b \left(P^a(t)+P^b(t)\right)/2 $). Using a
generic hidden order strategy $h=(h^a,h^b)\in\mathscr{H}$, the
trader's stock \textbf{inventory} and \textbf{cash} amount from
trading hidden orders are
\begin{equation}\label{terminal hid}
\begin{split}
I^h(t) = & \Delta^a \int_0^t h^a(s-)dH^a(s) - \Delta^b \int_0^t h^b(s-)dH^b(s);\\
C^h(t) = & -\Delta^a \int_0^t h^a(s-)
\left(\left(P^a(t)+P^b(t)\right)/2\right) dH^a(s)\\
& + \Delta^b \int_0^t h^b(s-)
\left(\left(P^a(t)+P^b(t)\right)/2\right)
 dH^b(s)\text{, }0\leq t\leq T.
\end{split}
\end{equation}

\section{The optimal trading problem}\label{sec trading}
Suppose the collective activities of the market participants form
the order book dynamics are described in the previous section. The
optimal \textit{trader} will trade on top of this aggregated
dynamics. He places a combination of active and hidden orders. The
active orders will immediately ``internalize" incoming market
orders. His hidden orders in the dark pool may or may not be
executed at the next moment, but once they are executed the trader
receives a price half the spread lower or higher than the current
ask or bid price. The flexibility to choose between active and
hidden orders enables finding an optimal balance between taking the
decision to internalize orders providing them a price improvement,
and the naturally associated adverse selection he is
exposed to via his inventory.\\
\\Depending on his informational advantage, the trader is identified
as either \textit{regular} or \textit{internalizing}. Most traders
in today's markets, including all the traders in Europe, are
regular. They observe and only observe the current records in the
order book. An internalizing trader has the priority of observing
incoming orders and acting upon them immediately before the orders
are displayed to other market participants. It offers an additional
choice to buy or sell at a slightly inferior price so that there is
no impact on the current best price.

\subsection{Actively filling displayed orders}\label{subsec trading
active}

An active trading strategy places orders that will be fully and
immediately executed at the best prices available, minus a ``price
improvement" offered to his counterpart if necessary.
 %---When new orders are arriving
%throughout a pipeline, the trader has the privilege of watching
%them, placing his own orders simultaneously and getting his orders
%executed ahead of those from the noise traders. Because large
%trading firms are usually very well equipped with high-quality
%hardware and are located close to the exchange, their priority in
%the execution is reasonable. The privilege to observe extra
%information is part of a VIP membership in the exchange.
\\
\\The total shares that the trader has
bought and sold up till time $t$ by actively internalizing the
market orders consist of those filled at the best available price,
denoted by the increasing processes $\{Z^a(t)\}_{0\leq t\leq T}$ and
$\{Z^b(t)\}_{0\leq t\leq T}$, and of those filled at the old prices
plus/minus a price improvement $\epsilon>0$ when new limit orders
arrive within the spread, denoted by the processes
$\{\beta^a(t)\}_{0\leq t\leq T}$ and $\{\beta^b(t)\}_{0\leq t\leq
T}$ as the proportion of shares filled at the old price with respect
to the total number of existing shares at the old price right before
the transaction. The active trading strategy
$Z=(Z^a,Z^b,\beta^a,\beta^b)$ is
called\\

-- \textit{continuously re-balanced}, if $Z^a$ and $Z^b$ are
continuous in the time $t$;

-- \textit{discretely re-balanced}, if $Z^a$ and $Z^b$ are pure jump
processes;

-- \textit{mixed}, if $Z^a$ and $Z^b$ are mixtures of the above two.\\
\\This subsection formulates mixed active trading
strategies (``mixed trading strategies" for short) respectively for
an internalizing trader  and a regular trader.
\begin{defn}\label{adm str 1.1} (mixed trading strategy)\\
(1) The set of admissible mixed trading strategies
$\mathscr{Z}^{int}$ of an internalizing trader is the
collection of all trading strategies $Z=(Z^a,Z^b,\beta^a,\beta^b)$ satisfying the following two criteria.\\
(1.1) The $\filtr$-adapted c\`{a}dl\`{a}g processes
$\{Z^a(t)\}_{0\leq t\leq T}$ and $\{Z^b(t)\}_{0\leq t\leq T}$ are
non-negative and non-decreasing over the time interval $[0,T]$. The
$\filtr$-adapted processes $\{\beta^a(t)\}_{0\leq
t\leq T}$ and $\{\beta^b(t)\}_{0\leq t\leq T}$ take values within the interval $[0,1]$. \\
(1.2) For two given positive integers $\underline{p}^b < \bar{p}^a$,
the process $Z^a$ is flat on
\begin{equation}
\{(t,\omega)\in [0,T]\times\Omega |P^a(t)\geq \bar{p}^a\},
\end{equation}
and $Z^b$ is flat on
\begin{equation}
\{(t,\omega)\in [0,T]\times\Omega |P^b(t)\leq \underline{p}^b\};
\end{equation}
the process $\beta^a$ is non-zero only on
\begin{equation}
\{(t,\omega)\in [0,T]\times\Omega |P^a(t)< \bar{p}^a\text{,
}N^a(t)-N^a(t-)=1\text{ and }Z^a(t)-Z^a(t-)<\Delta^a\},
\end{equation}
and the process $\beta^b$ is non-zero only on
\begin{equation}
\{(t,\omega)\in [0,T]\times\Omega |P^b(t)> \underline{p}^b\text{,
}N^b(t)-N^b(t-)=1\text{ and }Z^b(t)-Z^b(t-)<\Delta^b\}.
\end{equation}
(2) The set of admissible mixed trading strategies
$\mathscr{Z}^{reg}$ of a regular trader is defined as
\begin{equation}\label{defn mix reg}
\mathscr{Z}^{reg}:=\left\{Z=(Z^a,Z^b,\beta^a,\beta^b)\in\mathscr{Z}^{int}\left|
\beta^a(t) \equiv \beta^b(t) \equiv 0\text{, for all
}t\in[0,T]\right. \right\}.
\end{equation}
%for some $\epsilon>0$. Because the $\epsilon$ premium applies only
%when $\beta^a(t)=0$ or $\beta^b(t)=0$, the right hand side of
%equation $(\refeq{defn mix reg})$ is the same for all $\epsilon>0$.
\end{defn}
Criterion \textit{(1.1)} in Definition \ref{adm str 1.1} defines an
internalizing trader's mixed trading strategy. The bounded prices in
Criterion \textit{(1.2)} requires that the trader only buy below the
price $\bar{p}^a$ and sell above the price $\underline{p}^b$. In
practice, when the prices goes outside of their normal range, no
trading reflects a psyche of not to take up risk or push the prices
further
towards the extreme; technically, %Criterion \textit{(2)}
it will be used to prove the well-posedness of Problem \ref{prob
switch}, or in other words that the value function is finite. The
other requirement about $\beta^a$ and $\beta^b$ in Criterion
\textit{(1.2)} means that it is only necessary to consider filling a
fractional column at the old price when new limit orders arrive
within the spread. Allowing for optimizing over this fraction
enables a path-wise replication of the effect of every mixed trading
strategy by a price switching strategy, as will be shown in section
\ref{sec equiv}. A regular trader's mixed trading strategy defined
in Criterion \textit{(2)} is the same as that of an internalizing
trader, except that at the time of order arrival within the spread,
the regular trader can no longer fill the orders at the
old price, which is formulated as $\beta^a \equiv \beta^b \equiv 0$.\\
\\ The rest of the current subsection will write, in the
case of an internalizing trader, the order book dynamics and the
trader's stock inventory and cash amount in compact formulae. They
can be verified by enumerating all the situations that could trigger
a change in the bid or ask price. To adjust to the case of a regular
trader is only
a matter of setting $\beta^a \equiv \beta^b \equiv 0$.\\
\\The order book dynamics from equations (\refeq{price 0.3}),
(\refeq{price 0.4}) and (\refeq{price 0.2}), is now controlled by
the trader using an admissible mixed trading strategy $Z$. The
number of times over $[0,t]$ that all the orders at the current ask
and bid prices are depleted can be expressed as
\begin{equation}\label{price 1.3}
L^i(t)= \sum\limits_{0\leq s\leq t}\ind{Q^i(s-) - (Z^i(s) - Z^i(s-))
\leq 0}\text{, }i=a,b.
\end{equation}
For $i=a,b$, the changes
\begin{equation}\label{price 1.5}
\mu^a(t)= P^a(t)-P^a(t-)\text{ and }\mu^b(t)=
-\left(P^b(t)-P^b(t-)\right)
\end{equation}
 in the controlled ask and bid price processes at time $t$ can be
 computed by
\begin{equation}\label{price1.5}
\mu^i(t) = \left\{ \begin{aligned} -&1\text{, if }N^i(t)-N^i(t-)=1\text{ and }Z^i(t)-Z^i(t-)<\Delta^i;\\
  & 0 \text{, if } N^i(t)-N^i(t-)=1\text{ and }\Delta^i\leq Z^i(t)-Z^i(t-)<Q^i(t-)+\Delta^i,\\
 &~~~ \text{ or if } t=T\text{ and } Z^i(T)-Z^i(T-)<Q^i(T-);\\
&\left \lfloor \left(Z^i(t) - Z^i(t-) - Q^i(t-)\right) / \Delta^i
\right \rfloor +1 - (N^i(t)-N^i(t-))\text{, else}.
 \end{aligned} \right.
\end{equation}
The controlled \textbf{volumes} $Q^a$ and $Q^b$ at the ask and bid
prices move according to
\begin{equation}\label{price 1.2}
\begin{split}
Q^i(t)= & Q^i_0 + \sigma^i W^i(t) -Z^i(t) + \Delta^i
 \int_0^t(\mu^i(s))^+ dL^i(s)\\
& + \int_0^t \ind{\mu^i(s)\leq 0}\left(\Delta^i-(\mu^i(s))^-Q^i(s-)
\right)d N^i(s)\text{, }0\leq t\leq T\text{, }i=a,b.
\end{split}
\end{equation}
The controlled ask and bid \textbf{prices} $P^a$ and $P^b$ move
according to
\begin{equation}\label{price 1.4}
\begin{split}
P^a(t)= & P^a(0)+ \int_0^t (\mu^a(s))^+ dL^a(s)-\int_0^t
(\mu^a(s))^-dN^a(s);\\
P^b(t)= & P^b(0)-\int_0^t (\mu^b(s))^+ dL^b(s)+\int_0^t
(\mu^b(s))^-dN^b(s)\text{, }0\leq t\leq T.
\end{split}
\end{equation}
At every time $t\in [0,T]$, a mixed trading strategy $Z$ gives the
trader an \textbf{inventory} of stock shares
\begin{equation}\label{inventory 3.1}
I^Z(t) =  Z^a(t) - Z^b(t)+\int_0^t \beta^a(s)Q^a(s-)dN^a(s)
-\int_0^t \beta^b(s)Q^b(s-)dN^b(s).
\end{equation}
Let $\epsilon$ be the premium per share that the trader pays his
counterpart, for internalizing limit orders at the old price upon
the arrival of incoming orders at the new better price. Using the
mixed trading strategy $Z$, the total cash amount that the trader
pays the seller (receives from the buyer) of the stock denoted as
$C^Z_a$ (respectively $C^Z_b$) can be calculated by
\begin{equation}\label{cash 3.1.1}
\begin{split}
C^Z_a(t) = & \int_0^t P^a(s-)dZ^a(s) +\int_0^t
\beta^a(s)Q^a(s-)\left(P^a(s-)+\epsilon\right)dN^a(s)\\
& + \int_0^t \left( \frac{1}{2}\Delta^a d(P^a(s))^2
 +\left( \frac{1}{2}\Delta^a -Q^a(s) \right)dP^a(s)-\Delta^a
 dN^a(s)\right),
\end{split}
\end{equation}
and
\begin{equation}\label{cash 3.1.2}
\begin{split}
C^Z_b(t) = & \int_0^t P^b(s-)dZ^b(s)+\int_0^t
\beta^b(s)Q^b(s-)\left(P^b(s-)-\epsilon\right)dN^b(s)\\ & - \int_0^t
\left( \frac{1}{2}\Delta^b d(P^b(s))^2
 +\left( \frac{1}{2}\Delta^b -Q^b(s) \right)dP^b(s)-\Delta^b
 dN^b(s)\right).
\end{split}
\end{equation}
In the equation (\refeq{inventory 3.1}), the two integrals count the
total number of shares bought and sold at the old prices when limit
orders arrive within the spread. In each of the equations
(\refeq{cash 3.1.1}) and (\refeq{cash 3.1.2}), the first integral on
the right hand side of the identity is the amount of cash paid and
received if all the orders placed at the best available prices were
executed at the bid and ask prices observed right before each
transaction, as they are when the prices do not change and there is
no arrival within the spread. The second integral counts the cash
amount paid and received from filling the limit orders at the old
prices when new limit orders arrive within the spread. The third
integral collects the additional cost paid for trading deeper into
the book and the savings from filling the arriving limit orders
within the spread.\\
\\%Suppose the positive number $\epsilon$ is the premium that the
%trading platform pays the trader for actively buying or selling
%every one unit currency worth of the displayed limit orders,
The trader's cumulative \textbf{cash} amount on his account at time
$t$ from the mixed trading strategy $Z$ is
\begin{equation}\label{cash 3.1}
C^Z(t) = -C^Z_a(t) + C^Z_b(t)\text{, }0\leq t\leq T.
\end{equation}

\subsection{The goal of trading}

The trader's total stock \textbf{inventory} and \textbf{cash} amount
are sums of those from trading hidden and displayed orders, being
\begin{equation}\label{invent cash 1.1}
I^{h,Z}(t)=I_0+I^h(t)+I^Z(t)\text{ and
}C^{h,Z}(t)=C_0+C^h(t)+C^Z(t)\text{, }0\leq t\leq T,
\end{equation}
where the real numbers $I_0$ and $C_0$ are the initial stock
inventory and the initial cash amount, and the processes $I^h$,
$C^h$, $I^Z$ and $C^Z$ are defined in the equations (\refeq{terminal
hid}), (\refeq{inventory 3.1}) and (\refeq{cash 3.1}). When there is
no ambiguity on which trading strategies are
used, the superscripts in the stock inventory $I^{h,Z}$ and the cash amount $C^{h,Z}$ are omitted.\\
\\Let $r^I\in (-\infty,\infty)$ and $r^C\in (0,\infty)$ be two real numbers and
\begin{equation}\label{rwd 1.2}
F:\re\times \mathbb{N}\times \mathbb{N}\rightarrow \re \text{; }(z,p^a,p^b)\mapsto F(z,p^a,p^b),
\end{equation}
be a measurable function with  quadratic growth.
 \begin{assump}\label{assump rwd}
 There exists a constant $r^F>0$, such that for all
 $z \in \re$ and all $ p^b\leq p^a$, we have
 \begin{equation}
 \left|F(z,p^a,p^b)\right| \leq r^F \left((z)^2+(p^a)^2+(p^b)^2+1\right)
  \end{equation}
and
 \begin{equation}
 \left|F(z_1,p^a,p^b)-F(z_2,p^a,p^b)\right| \leq r^F \left(|p^a|+|p^b|+1\right)
\left(|z_1|+|z_2|+1\right)|z_1-z_2|.
  \end{equation}

 \end{assump}
The trading activities are measured by the
 reward
 \begin{equation}\label{rwd 1.3}
 \xi\left(I(T),C(T)\right)= r^C C(T) + r^I F\left(I(T),P^a(T),P^b(T)\right).
 \end{equation}
 The trader's objective of maximizing the reward in expectation is
 formulated as a stochastic control problem.
\begin{problem}\label{prob sing}
(1) An internalizing trader  looks for an optimal trading strategy\\
$Z^*=(Z^{a*},Z^{b*}, \beta^{a*},\beta^{b*})\in\mathscr{Z}^{int}$ and
$h^*=(h^{a*},h^{b*})\in\mathscr{H}$ to achieve the maximum expected
reward
\begin{equation}\label{exp rwd 1.2}
V^{mix,int}(\epsilon):=\sup\limits_{(Z^a,Z^b,\beta^a,\beta^b)\in\mathscr{Z}^{int},(h^a,h^b)\in\mathscr{H}}
\expect\left[\xi\left(I^{h,Z}(T),C^{h,Z}(T)\right)\right].
\end{equation}
The reward $V^{mix,int}(\epsilon)$ is a function of the premium
$\epsilon$, because the cash amount $C^{h,Z}(T)$ is a function of $\epsilon$.\\
(2) A regular trader looks for an optimal trading strategy
$Z^*=(Z^{a*},Z^{b*},0,0)\in\mathscr{Z}^{reg}$ and
$h^*=(h^{a*},h^{b*})\in\mathscr{H}$ to achieve the maximum expected
reward
\begin{equation}\label{exp rwd 1.3}
V^{mix,reg}:=\sup\limits_{(Z^a,Z^b,0,0)\in\mathscr{Z}^{reg},(h^a,h^b)\in\mathscr{H}}
\expect\left[\xi\left(I^{h,Z}(T),C^{h,Z}(T)\right)\right].
\end{equation}
\end{problem}
%The part on choosing an optimal $Z^*$ is a mixed control problem.
Control problems with reward functions of the form (\refeq{rwd 1.3})
have a five dimensional state process $(Q^a,Q^b,I,P^a,P^b)$. Problem
\ref{prob sing} is still solvable for reward functions that are not
linear in $C(T)$, in which case the state process will have the cash
amount $C$ a sixth dimension. Assumption \ref{assump rwd} is a
technical assumption under which Problem \ref{prob switch} and thus
Problem \ref{prob sing} are well-posed.\\
\\Several common situations where the reward criteria satisfy
Assumption \ref{assump rwd} are linear combination of the cash and
inventory, liquidating or filling a certain number of stock shares,
and holding cash only at the terminal time,
 corresponding to the following forms of $\xi(I(T),C(T))$ defined in (\refeq{rwd 1.3}):
  \begin{equation}\label{rwd 1.1}
\begin{split}
\textit{(1) } &\xi(I(T),C(T))= r^C C(T) + r^I I(T);\\
\textit{(2) } &\xi(I(T),C(T))= r^C C(T) + r^I |{I(T)}-z_0|\text{ or }\xi(I(T),C(T))= r^C C(T) + r^I ({I(T)}-z_0)^2; \\
\textit{(3) } &\xi(I(T),C(T))= r^C C(T) + r^I
\left(\ind{I(T)>0}(P^b(T)-U^b)+\ind{I(T)<0}(P^a(T)+U^a)\right)I(T).
\end{split}
\end{equation}
The criterion (\refeq{rwd 1.1})\textit{(1)} means that the trader
has a utility function linear in cash and inventory. In (\refeq{rwd
1.1})\textit{(2)}, the coefficient $r^I$ is negative, and the
constant $z_0$ is the number of shares that the trader would like to
hold at the terminal time $T$. In (\refeq{rwd 1.1})\textit{(3)}, the
coefficient $r^I$ is positive and $U^a$ and $U^b$ are two positive
integers; if the terminal inventory $I(T)$ is positive, the trader
sells all his stocks at the price $(P^b(T)-U^b)$ per share; if
$I(T)$ is negative, he pays the price $(P^a(T)+U^a)$ for each
share.\\
\begin{rmk}
In application, a trader in a hedge fund or a proprietary trading
firm is entitled to both buying and selling during any trading
period, hence $Z^a$, $Z^b$, $h^a$ and $h^b$ can be all positive. For
a trader in a brokerage agency, if he trades during the time $[0,T]$
to fill a buy (sell) order for the customer, a simplest way to
comply with the regulations is not to sell (buy) throughout the same
time period, hence, when using the results in this paper, he should
set $Z^b(t)\equiv 0$, $\beta^b(t)\equiv 0$ and $h^b(t)\equiv 0$
(respectively $Z^a(t)\equiv 0$, $\beta^a(t)\equiv 0$ and
$h^a(t)\equiv 0$) for all $0\leq t\leq T$. Especially, a regular
trader sets $\beta^a \equiv \beta^b\equiv 0$. The brokerage agency's
admissible set of equivalent price switching strategies is a
modification of Definition \ref{adm str 2.1} by setting $u^b_n
\equiv -(N^b(S_n)-N^b(S_n-))$ and $h^b(t)\equiv 0$ (respectively
$u^a_n \equiv - (N^a(S_n)-N^a(S_n-))$ and $h^a(t)\equiv 0$).
Optimizing over the buying (selling) strategies only is a special
case of the algorithm in subsection \ref{subsec numeric}.
\end{rmk}

\section{The optimal switching problem}\label{sec switching}

This section presents the problem of how the trader could optimally
switch the bid and ask prices and shows the well-posedness of the
optimal switching problem. Compared to the optimal trading Problem
\ref{prob sing}, price switching considers a smaller and simpler set
of active trading strategies, which are discretely re-balanced
strategies in the form of the times to trade and the number of
limits to fill at each time.
%The equivalence of the two problems will be proven in the
%next section.

\subsection{Switching prices}\label{subsec switching
active} Let $[u]$ and $\{u\}$ respectively denote the integer part
and the fractional part of a real number $u$. The identity
$u=[u]+\{u\}$ holds. Switching the prices means that the trader
chooses a sequence of times $\{S_n\}_{n=1}^\infty$ and two sequences
of positive numbers $\{u^a_n\}_{n=1}^\infty$ and
$\{u^b_n\}_{n=1}^\infty$, such that the ask and bid prices are
pushed from $P^a(S_{n-1})$ to $P^a(S_{n})=P^a(S_{n-1})+[u^a_n]$ and
from $P^b(S_{n-1})$ to $P^b(S_{n})=P^b(S_{n-1})-[u^b_n]$. The ask
and bid prices stay
constants over every time interval $(S_{n-1},S_n)$.\\
\\In the limit order book where the trader and the rest of the market participants all
act, a price change could occur due to two possible reasons.\\
\\(1) Limit orders at the ask or bid price is depleted by the noise
trader, or limit orders arrive within the spread. Suppose the
previous time of price switching is $S_{n-1}$, then the next time
$T_n$ when either event happens can be expressed as
\begin{equation}\label{time change}
\begin{split}
T_n:=\inf\{&S_{n-1}<t\leq T |
Q^a(S_{n-1})+\sigma^a(W^a(t)-W^a(S_{n-1}))= 0\text{,
}N^a(t)-N^a(S_{n-1})=1,\\
&Q^b(S_{n-1})+\sigma^b(W^b(t)-W^b(S_{n-1}))= 0\text{, or
}N^b(t)-N^b(S_{n-1})=1 \}\wedge T.
\end{split}
\end{equation}
(2) The trader fills all the shares at the ask or bid price and, at
his choice, some limits beyond the best prices. If he trades at some
time $S_n\in (S_{n-1}, T_{n})$, then he has to fill all the shares
at either the ask price or the bid price to trigger a price
change.\\
\\After the $(n-1)$th price switching at the time $S_{n-1}$, if he
waits  until the time $T_n$, the trader may choose to fill some
shares at time $T_n$; even if he does not trade, the rest of the
market will switch the prices at time $T_n$ any way, in which case
the time of the $n$th price switching is set as $S_n=T_n$. Requiring
that the trader has to ``trade" at time $T_n$, though possibly zero
share, makes sure that the prices are constants between two
switching times and gives a
neater expression of the order book dynamics.\\
\\Furthermore, formulating  with the help of $T_n$'s
reduces the price switching problem from seven-dimensional
path-dependent to five-dimensional Markovian. The actual controlled
state process is
\begin{equation}
\{(Q^a(t-), Q^b(t-), I^\alpha(t-)+I^h(t), P^a(t-),P^b(t-),
N^a(t),N^b(t))\}_{0\leq t\leq T}.
\end{equation}
Since what matters in  $(N^a(t),N^b(t))$ is their exponentially
distributed arrival times, the trader only need to monitor whether
or not there is an arrival of limit sell or buy orders within the
spread. Viewing the controlled arrival times $T_n$'s as a sequence
of exit times when decision marking has to take place, the state
process is simplified to $\{(Q^a(t-), Q^b(t-), I^\alpha(t-)+I^h(t),
P^a(t-),P^b(t-))\}_{0\leq t\leq T}$.
\\
\\The admissible price switching strategies are switching controls defined below.
\begin{defn}\label{adm str 2.1} (switching control) The admissible set of
switching controls of an internalizing trader  and that of a regular
trader are denoted respectively as $\mathscr{A}^{int}$ and
$\mathscr{A}^{reg}$. Let the letter $j$ represent either the
superscripts ``int" or the superscript ``reg". The admissible set
$\mathscr{A}^j$ consists of switching controls
$\alpha:=\{(S_n,u_n^a,u_n^b)\}_{n=0}^\infty$ satisfying, for every
$n=1,2,\cdots$, the three criteria below, with the convention
that $S_0=u^a_0=u^b_0=0$.\\
(1) The switching time $S_{n}$ is an $\filtr$-stopping time such
that $0 \leq S_1 < \cdots < S_{n-1}< S_{n}<\cdots$. If $S_{n-1}\geq
T$, then $S_n=T+1$; if $S_{n-1} < T$, then $S_n\in (S_{n-1},
T_{n}]$, where $T_n$ defined in (\refeq{time change}) is the next
time of price change if the trader does not trade.\\ (2) For the
same positive integers $\bar{p}^a$ and $\underline{p}^b$ as in
Definition \ref{adm str 1.1}, we have $(u^a_n,u^b_n)\in
\mathscr{U}^j(S_n,P^a(S_{n-1}),P^b(S_{n-1}))$, where
\begin{equation}\label{adm switch}
\mathscr{U}^j(t,p^a,p^b):=\left\{\begin{aligned}
&\{0,1,\cdots,(\bar{p}^a-p^a)^+ \}\times
\{0,1,\cdots,(p^b-\underline{p}^b)^+ \}\setminus
\{(0,0)\},\\
&~~~~\text{if } t<T_n\text{, or }
Q^i(t-)=0 \text{, } i=a\text{ or }b;\\
&D^j(\bar{p}^a-p^a)\cup \{2,\cdots,(\bar{p}^a-p^a)^+ \}\times
\{0,1,\cdots,(p^b-\underline{p}^b)^+ \},\\
&~~~~~~~~~~~~~~~~~~~~\text{if }N^a(t)-N^a(t-)=1;\\
&\{0,1,\cdots,(\bar{p}^a-p^a)^+ \}\times
D^j(p^b-\underline{p}^b)\cup \{2,\cdots,(p^b-\underline{p}^b)^+
\},\\&~~~~~~~~~~~~~~~~~~~~~\text{if
}N^b(t)-N^b(t-)=1;\\
%&\{0,1,\cdots,(\bar{p}^a-p^a)^+ \}\times
%\{0,1,\cdots,(p^b-\underline{p}^b)^+ \}\text{, if } Q^a(t-)=0 \text{ or } Q^b(t-)=0;\\
& [0,(\bar{p}^a-p^a)^+ ]\times [0,(p^b-\underline{p}^b)^+ ]\text{, if } t=T; \\
&\{(0,0)\}\text{, if } t=T+1.
\end{aligned}\right.
\end{equation}
In expression (\refeq{adm switch}), the sets $D^j(\bar{p}^a-p^a)$
and $D^j(p^b-\underline{p}^b)$ of real numbers are defined as
\begin{equation}\label{set Dp}
D^{int}(x):=\left\{\begin{aligned}
&[-1,1]\text{, if }x \geq 1;\\
  &\{-1,0\}\text{, if }x =0;\\
   &\{-1\}\text{, if }x \leq -1
   \end{aligned}\right.
\end{equation}
for an internalizing trader, and as
\begin{equation}\label{set Dr}
D^{reg}(x):=\left\{\begin{aligned}
&\{-1,0,1\}\text{, if }x \geq 1;\\
  &\{-1,0\}\text{, if }x =0;\\
   &\{-1\}\text{, if }x \leq -1.
   \end{aligned}\right.
\end{equation}
for a regular trader.
 For $j=int,reg$ and $t\in [0,T]$, the set $\mathscr{A}^j_t$ is defined
the same as $\mathscr{A}^j$ except that $S_0=t$. Also, we denote by
\begin{equation}
\mathscr{A}^j_{t,1}:=\left\{ \left (S_1,u_1^a,u_1^b) \right|
\alpha=\{(S_n,u_n^a,u_n^b)\}_{n=0}^\infty \in \mathscr{A}^j_t
\right\}
\end{equation}
the set of the first elements of all switching controls in
$\mathscr{A}^j_t$.
\end{defn}
In Definition \ref{adm str 2.1}, Criterion \textit{(1)} specifies
the trading times $S_n$. Criterion \textit{(2)} specifies the
possible numbers of limits to buy and sell at each transaction.
Setting $S_n=T+1$ means the trader no longer trades, hence there has
to be $u^a_n=u^b_n= 0$.\\
\\The admissible switching control sets of an internalizing trader  and a
regular trader are different only up to the sets $D^{int}$ and
$D^{reg}$ defined in (\refeq{set Dp}) and (\refeq{set Dr}).
Throughout this paper, when a claim is valid for a trader regardless
of whether he is internalizing or regular, the admissible switching
control sets will be generically
denoted as $\mathscr{A}$, $\mathscr{A}_t$ and $\mathscr{A}_{t,1}$.\\
\\For $n=1,2,\cdots$, at every
time $S_n$ a switching control $\alpha\in \mathscr{A}$ causes the
ask price to increase from $P^a(S_{n-1})$ to
$P^a(S_{n})=P^a(S_{n-1})+[u^a_n]$ and the bid  price to decrease
from $P^b(S_{n-1})$ to $P^b(S_{n})=P^b(S_{n-1})-[u^b_n]$. By
enumerating all the situations that could trigger a price change,
the order book dynamics and the changes in the stock
inventory and cash amount can be summarized in compact formulae.\\
\\The order book dynamics can be written in terms of the switching
control
\begin{equation}
\alpha=\{(S_n,u_n^a,u_n^b)\}_{n=0}^\infty \in \mathscr{A}:
\end{equation}
the controlled \textbf{prices} move according to
\begin{equation}\label{price 2.4}
P^a(t)=P^a(0)+\sum\limits_{n:S_n\leq t} [u^a_n] \text{, and }
P^b(t)=P^b(0)-\sum\limits_{n:S_n\leq t}[u^b_n] \text{, }0\leq t\leq
T;
\end{equation}
the controlled \textbf{volumes} move according to
\begin{equation}\label{price 2.2}
%\begin{split}
Q^i(t) =  \left\{
\begin{aligned}& Q^i(0) + \sigma^i W^i(t)\\
& + \sum\limits_{n:S_n\leq t} \left( \ind{[u^i_n]\neq
0}\left(\Delta^i-Q^i(S_n-)\right)-\ind{[u^i_n]= 0}\{u^i_n\}Q^i(S_n-)
\right)\text{, }0\leq t< T\text{; }\\
& \sum\limits_{n:S_n = T} (1-\{u^i_n\})\left(\ind{u^i_n\geq 1}
\Delta^i +\ind{[u^i_n]= 0} Q^i(T-)\right)\text{, }t=T\text{, }i=a,b.
\end{aligned}
\right.
%\end{split}
\end{equation}
Defining the functions  $g^a$ and $g^b: \Omega\times [0,T]\times
[0,\infty)\times \mathbb{Z}\rightarrow \re$ as
\begin{equation}\label{rwd2.2.1}
g^i(t,q,u) = \left\{
\begin{aligned} &\ind{u\geq 1}\left(q +  (u-1) \Delta^i\right)
+ (N^i(t)-N^i(t-))\left(\ind{u\geq 0}\Delta^i + \ind{[u]\leq
0}\{u\} q\right)\text{, }t<T;\\
&\ind{u\geq 1}(q+ (u-1)\Delta^i) + \ind{[u]=0}u\cdot  q\text{,
}t=T\text{, }i=a,b,
\end{aligned}
\right.
\end{equation}
then $g^a\left(S_n, Q^a(S_{n}-), u^a_{n}\right)$ and $g^b\left(S_n,
Q^b(S_{n}-), u^b_{n}\right)$ are respectively the number of shares
that the trader buys and sells at the time $S_n$. The quantity
\begin{equation}
\begin{split}
g_\alpha\left(S_n, Q^a(S_{n}-), Q^b(S_{n}-), u^a_{n}, u^b_{n}\right)
:=  g^a\left(S_n, Q^a(S_{n}-),  u^a_{n}\right) - g^b\left(S_n,
Q^b(S_{n}-), u^b_{n}\right)
\end{split}
\end{equation}
is the change in the trader's stock inventory from
transactions at the time $S_n$.\\
\\Let $\epsilon$ be the premium that the trader pays his counterpart for
filling each share at the old price upon the arrival of limit orders
at the new price. Defining the functions $f^a$ and $f^b:
\Omega\times [0,T]\times [0,\infty)\times \mathbb{Z} \times
\mathbb{N}\rightarrow \re$ as
\begin{equation}\label{rwd2.2.2}
\begin{split}
&f^a(t,q,p,u) = \left\{
\begin{aligned}
&\ind{u\geq 1}\left(p\left(q+ (u-1)\Delta^a\right) +
\frac{1}{2}u(u-1)\Delta^a\right) \\&+
(N^a(t)-N^a(t-))\left(\ind{u\geq 0}(p-1)\Delta^a+\ind{[u]\leq
0}(p+\epsilon)\{u\}q\right)\text{, }t<T;\\
&\ind{u\geq 1}\left(p\left(q+
(u-1)\Delta^a\right)+\left(\frac{1}{2}[u]([u]-1)+[u]\cdot\{u\}\right)\Delta^a\right)
\\
&+\ind{[u]=0}p\cdot u\cdot q\text{, }t=T,
\end{aligned}
\right.
\end{split}
\end{equation}
and
\begin{equation}\label{rwd2.2.3}
\begin{split}
&f^b(t,q,p,u) = \left\{
\begin{aligned}
&\ind{u\geq 1}\left(p\left(q+ (u-1)\Delta^b\right) +
\frac{1}{2}u(u-1)\Delta^b\right) \\&+
(N^b(t)-N^b(t-))\left(\ind{u\geq 0}(p+1)\Delta^b+\ind{[u]\leq
0}(p-\epsilon)\{u\}q\right)\text{, }t<T;\\
&\ind{u\geq 1}\left(p\left(q+ (u-1)\Delta^b\right) +
\left(\frac{1}{2}[u]([u]-1)+[u]\cdot\{u\}\right)\Delta^b\right)\\
&+\ind{[u]=0}p\cdot u\cdot q\text{, }t=T,
\end{aligned}
\right.
\end{split}
\end{equation}
then $f^a\left(S_n, Q^a(S_{n}-), P^a(S_{n}-), u^a_{n}\right)$ and
$f^b\left(S_n, Q^b(S_{n}-), P^b(S_{n}-), u^b_{n}\right)$ are
respectively the total cash amount that the trader pays the seller
(receives from the buyer) of the stock for transactions at the time
$S_n$. %Adjusting for the  $\epsilon$ premium paid by the trading
%platform for actively buying or selling every one unit currency
%worth,
The quantity
\begin{equation}\label{rwd2.2.4}
\begin{split}
&f_\alpha\left(S_n, Q^a(S_{n}-), Q^b(S_{n}-), P^a(S_{n}-),
P^b(S_{n}-), u^a_{n}, u^b_{n}\right)\\
: = & - f^a\left(S_n, Q^a(S_{n}-), P^a(S_{n}-), u^a_{n}\right) +
f^b\left(S_n, Q^b(S_{n}-), P^b(S_{n}-), u^b_{n}\right)
\end{split}
\end{equation}
is the change in the trader's cash amount on his account from
transactions at the
time $S_n$.\\
\\Actively trading according to a generic switching control
${\alpha}\in\mathscr{A}$ defined by Definition \ref{adm str 2.1},
the trader's \textbf{inventory} and \textbf{cash} amount from the
displayed orders are respectively
\begin{equation}\label{rwd2.1}
\begin{split}
I^{\alpha}(t) = &\sum\limits_{n:S_n\leq t} \, g_\alpha\left(S_n,
Q^a(S_{n}-), Q^b(S_{n}-), u^a_{n}, u^b_{n}\right)\text{ and }\\
C^{\alpha}(t) %= &\sum\limits_{n:S_n\leq t} \, f_\alpha\left(S_n,
%Q^a(S_{n}-), Q^b(S_{n}-),P^a(S_{n-1}),P^b(S_{n-1}), u^a_{n}, u^b_{n}\right)\\
=&\sum\limits_{n:S_n\leq t} \, f_\alpha\left(S_n, Q^a(S_{n}-),
Q^b(S_{n}-), P^a(S_{n}-),P^b(S_{n}-), u^a_{n}, u^b_{n}\right)\text{,
for }0\leq t\leq T.
\end{split}
\end{equation}
Taking into account the hidden orders, the trader's total terminal
stock \textbf{inventory} and \textbf{cash} amount are respectively
\begin{equation}\label{invent cash 2.1}
I(T)^{h,\alpha}=I_0+I^h(T)+I^{\alpha}(T)\text{ and
}C^{h,\alpha}(T)=C_0+C^h(T)+C^{\alpha}(T),
\end{equation}
where the quantities $I^h(T)$ and $C^h(T)$ are defined in the
equations (\refeq{terminal hid}). When there is no ambiguity on
which trading strategies are used, the superscripts in
$I^{h,\alpha}(T)$
and $C^{h,\alpha}(T)$ are omitted.\\
\\Since each transaction using active orders causes a change in the prices,
the trader's trading activities are a matter of choosing when and to
what level to ``switch" the ask and bid prices. The profit or cost
from switching at time $S_n$ is the change in his cash amount
expressed as the quantity in (\refeq{rwd2.2.4}). This is why we give
the name ``price switching" to the set of discretely re-balances
trading strategies introduced in this section. Optimizing the
trading algorithm over all the price switching strategies is a
problem of switching control with impact on the state process.
\begin{problem}\label{prob switch}
(1) An internalizing trader  looks for an admissible switching
control $\alpha^*\in\mathscr{A}^{int}$ and an optimal hidden order
strategy $h^*\in\mathscr{H} $ that achieve the supremum
\begin{equation}\label{goal2.3}
V^{swt,int}(\epsilon):=\sup\limits_{\alpha\in\mathscr{A}^{int},\,h\in\mathscr{H}}\expect[\xi(I(T)^{h,\alpha},C^{h,\alpha}(T))].
\end{equation}
Same as in Problem \ref{prob sing}, the reward
$V^{swt,int}(\epsilon)$ is a function of the premium $\epsilon$.\\
(2) A regular trader looks for an admissible switching control
$\alpha^*\in\mathscr{A}^{reg}$ and an optimal hidden order strategy
$h^*\in\mathscr{H} $ that achieve the supremum
\begin{equation}\label{goal2.6}
V^{swt,reg}:=\sup\limits_{\alpha\in\mathscr{A}^{reg},\,h\in\mathscr{H}}\expect[\xi(I(T)^{h,\alpha},C^{h,\alpha}(T))].
\end{equation}
\end{problem}
Generically for either an internalizing trader  or a regular trader,
Problem \ref{prob switch} requires finding an admissible switching
control $\alpha^*\in\mathscr{A}$ and an optimal hidden order
strategy $h^*\in\mathscr{H} $ that achieve the supremum
\begin{equation}\label{goal2.7}
\begin{split}
&\sup\limits_{\alpha\in\mathscr{A},\,h\in\mathscr{H}}\expect[\xi(I(T)^{h,\alpha},C^{h,\alpha}(T))]\\
=&\sup\limits_{\alpha\in\mathscr{A},\,h\in\mathscr{H}}\expect\bigg[r^C
\sum\limits_{n = 1}^\infty
f_\alpha\left(S_n,Q^a(S_{n}-),Q^b(S_{n}-),P^a(S_{n}-),P^b(S_{n}-),
u^a_{n},u^b_{n}\right)\\
& ~~~~~~~~~~~~~~~~~+r^C C^h(T)+ r^I
F(I^{h,\alpha}(T),P^a(T),P^b(T))\bigg].
\end{split}
\end{equation}
The \textbf{value process} $V$ of (\refeq{goal2.7}) is defined as
\begin{equation}\label{goal2.2}
\begin{split}
V(t):%=&\sup\limits_{\alpha\in\mathscr{A}_t,\,h\in\mathscr{H}_t}\expect\left[\left.r^C
%\sum\limits_{n = 1}^\infty f\left(S_{n-1},S_n,Q(S_{n}-),P(S_{n}-),
%u_{n},h\right) + r^I F(I(T),P(T))\right|\filtr(t)\right]\\
=& \sup\limits_{\alpha\in\mathscr{A}_t,\,h\in\mathscr{H}_t}
\expect\left[\left.\xi(I(T),C(T))\right|\filtr(t)\right]-r^C C(t) .
\end{split}
\end{equation}
Then the best expected reward (\refeq{goal2.7}) can be written as
\begin{equation}\label{goal2.4}
\sup\limits_{\alpha\in\mathscr{A},\,h\in\mathscr{H}}\expect[\xi(I(T),C(T))]
=r^C C_0+V(0).
\end{equation}

\subsection{Well-posedness of the problem}\label{subsec
wellposed}\label{subsec 4.2}

Before looking for an optimal trading strategy that achieves the
supremum in (\refeq{goal2.7}), it is necessary to verify that the
optimal switching problem is well-defined. In other words, is the
value process $V$ in (\refeq{goal2.2}) finite? This subsection will
prove that the answer is yes. Later in Proposition \ref{prop bound},
we shall see that the optimal price switching problems provide lower
and upper bounds of the value functions of the optimal trading
problem, hence the optimal trading problem is consequently also
well-posed. By Theorem \ref{thm equiv}(6)(8) and Proposition
\ref{prop bound}, we shall see that
\begin{equation}
V^{swt,reg} \leq V^{swt,int}(\epsilon)\text{, }
V^{mix,int}(\epsilon) \text{ and } V^{mix,reg}\leq V^{swt,int}(0).
\end{equation}
It suffices to prove the well-posedness of the regular trader's
price switching problem and the internalizing trader's price
switching problem with zero premium. Then the well-posedness of
Problem \ref{prob sing} and Problem \ref{prob switch} all follows.\\
\\Throughout this subsection, the
premium $\epsilon$ equals zero.\\
\\The well-posedness is stated as Theorem \ref{lem bound 2.1} at the end of this subsection.
To prepare for the proof of the theorem, Lemma \ref{lem bound price}
and Lemma \ref{lem bound inven cash} will respectively provide
uniform $\mathbb{L}^2$ or $\mathbb{L}^1$ bounds of the prices
$(P^a(T),P^b(T))$, the stock inventory $I^{h,\alpha}(T)$ and the
cash amount $C^{h,\alpha}(T)$ for all admissible switching controls.
Because the reward criterion $\xi$ defined in (\refeq{rwd 1.3}) is a
function of $(P^a(T)$, $P^b(T))$, $I^{h,\alpha}(T)$ and
$C^{h,\alpha}(T)$, the two Lemmas lead to Theorem \ref{lem bound
2.1}. In addition, an interpretation worth noting of Lemma \ref{lem
bound price} is that the trader's trading activities will not push
the prices towards explosion.

\begin{lem}\label{lem bound price} There exists a constant $c_1>0$, such that for any admissible switching control
$\alpha\in\mathscr{A}$ defined by Definition \ref{adm str 2.1}, the
controlled ask and bid prices at any time $t\in [0,T]$ has the
$\mathbb{L}^2$ bounds
\begin{equation}\label{bound 4.1.0}
\begin{split}
\expect\left[ \left(P^a(t) \right)^2\right]
\leq & (P^a(0))^2 +(\bar{p}^a)^2 + c_1T\left(((\theta^{a*})^2 + 1)T + \theta^{a*} + 1\right);    \\
\expect\left[ \left(P^b(t) \right)^2\right] \leq & (P^b(0))^2
+(\underline{p}^b)^2 + c_1T\left(((\theta^{b*})^2 + 1)T +
\theta^{b*} + 1\right),
\end{split}
\end{equation}
where the constants $\theta^{a*}$ and $\theta^{b*}$ defined in
(\refeq{notn theta star}) are the maximum arrival rates of limit
orders within the spread.
\end{lem}
\textbf{Proof.} The total number of downward (upward) movements in
the ask (bid) price equals the number of times $N^i$  that limit
sell (buy) orders arrive within the spread. The total number of
upward (downward) movements in the ask (bid) price equals the number
of limits that the trader has filled plus the number of times when
the volume at the ask (bid) price is depleted by the rest of the
market. The total number of depletions at the ask (bid) price by the
rest of the market participants does not exceed a renewal process
$R^i$ with independent inter-arrival times identically distributed
as the leverage hitting time
\begin{equation}
\inf \left\{0 \leq t \leq T \left| \Delta^i + \sigma^i W^i(t)\leq 0
\right. \right \}\wedge T \text{, for } i=a,b.
\end{equation}
The highest ask (lowest bid) price happens when no limit sell (buy)
orders arrive within the spread and the trader fills all the limit
sell (buy) order below the price $\bar{p}^a$ (above the price
$\underline{p}^b$). The lowest ask (highest bid) price happens when
there is no depletion by the trader or the noise trader, and limit
sell (buy) orders arrive according to the largest possible
intensity. The prices have the upper and lower bounds
\begin{equation}\label{bound 4.1.1}
\begin{split}
& P^a(0) - N^a(T) \leq P^a(0) - N^a(t) \leq  P^a(t) \leq \bar{p}^a\vee P^a(0) + R^a(t) \leq \bar{p}^a\vee P^a(0) + R^a(T);\\
& \underline{p}^b\wedge P^b(0) - R^b(T)  \leq \underline{p}^b\wedge
P^b(0) - R^b(t) \leq P^b(t) \leq  P^b(0) + N^b(t) \leq P^b(0) +
N^b(T).
\end{split}
\end{equation}
Using the inequalities in (\refeq{bound 4.1.1}) and the bounds for
variances of stationary renewal processes derived in
\cite{Daley1978} Daley (1978), we get the inequalities in
(\refeq{bound 4.1.0}). \hfill $\Box$

\begin{lem}\label{lem bound inven cash} There exist positive constants $c_2$ and $c_3$, such
that for any admissible switching control $\alpha\in\mathscr{A}$
defined by Definition \ref{adm str 2.1} and any hidden order
strategy $h\in\mathscr{H}$ defined by Definition \ref{adm str 1.2},
the trader's total cash amount and stock inventory from the
strategies $\alpha$ and $h$ have the bounds
\begin{equation}\label{est 1.4}
\expect[|C^h(T)+C^\alpha(T)|]\leq c_2
\left((Q^a(0))^2+(Q^b(0))^2+(P^a(0))^2+(P^b(0))^2+T^2+1\right)
\end{equation}
and
\begin{equation}\label{est 1.5}
\begin{split}
&\expect[|I_0+I^h(T)+I^\alpha(T)|^2]\\
\leq & c_3
\left((Q^a(0))^4+(Q^b(0))^4+(I_0)^2+(P^a(0))^4+(P^b(0))^4+T^4+1\right).
\end{split}
\end{equation}
\end{lem}
\textbf{Proof.} Throughout the time interval $[0,T]$, the ask price
$P^a$ (bid price $P^b$) moves downward  $N^a(T)$ times (upward
$N^b(T)$ times). At least one limit at a time, the trader could take
at most all the initially existing displayed limit sell (buy) orders
 and all the displayed limit sell (buy) orders ever arriving within the spread, as
 long as the ask (bid) price is below $\bar{p}^a$ (above $\underline{p}^b$).
 Then the total number of limits that the trader would ever buy (sell) until time $T$
  does not exceed $(\bar{p}^a-P^a(0))^+ + N^a(T)$
  (respectively $(P^b(0)-\underline{p}^b)^+ + N^b(T)$). At any time $t\in [0,T]$,
  each limit of limit sell (buy) orders contains no more than
  $\Delta^a + Q^a(0) + 2\sigma^a\sup\limits_{0\leq t\leq T}|W^a(t)|$
  (respectively $\Delta^b + Q^b(0) + 2\sigma^b\sup\limits_{0\leq t\leq T}|W^b(t)|$) shares.
  Each time buying (selling), it is impossible to take more than the number of all
  the currently existing limit sell (buy) orders below (above) the  price $\bar{p}^a$
  (respectively $\underline{p}^b$). We may bound the stock inventory and  cash amount
  from displayed orders by
 \begin{equation}\label{est 1.1}
\begin{split}
  |I^\alpha(T)|\leq &\left((\bar{p}^a-P^a(0))^+ + N^a(T)\right)
\left(\Delta^a + Q^a(0) + 2\sigma^a\sup\limits_{0\leq t\leq T}|W^a(t)|\right)\\
&+ \left((P^b(0)-\underline{p}^b)^+ + N^b(T)\right) \left(\Delta^b +
Q^b(0)
+ 2\sigma^b\sup\limits_{0\leq t\leq T}|W^b(t)|\right);\\
  |C^\alpha(T)|\leq & \left(\bar{p}^a+\epsilon\right)\left((\bar{p}^a-P^a(0))^+ + N^a(T)\right)
\left(\Delta^a + Q^a(0) + 2\sigma^a\sup\limits_{0\leq t\leq T}|W^a(t)|\right)\\
&+
\left(\underline{p}^b+\epsilon\right)\left((P^b(0)-\underline{p}^b)^+
+ N^b(T)\right) \left(\Delta^b + Q^b(0) +
2\sigma^b\sup\limits_{0\leq t\leq T}|W^b(t)|\right).
\end{split}
\end{equation}
Considering the hidden orders, the trader could receive at most
$\Delta^a H^a(T)$ shares of hidden limit sell orders and $\Delta^b
H^b(T)$ shares of hidden limit buy orders. The greatest possible
price for each share does not exceed the bounds in (\refeq{bound
4.1.1}). The stock inventory and cash amount from hidden orders can
be bounded by
 \begin{equation}\label{est 1.2}
\begin{split}
  |I^h(T)|\leq & \Delta^a H^a(T) + \Delta^b H^b(T);\\
  |C^h(T)|\leq & \Delta^a H^a(T)\left(P^a(0) + \bar{p}^a + N^a(T) + R^a(T)+1\right)\\
  & + \Delta^b
  H^b(T)\left(P^b(0) + \underline{p}^b + N^b(T) + R^b(T)+1\right).
\end{split}
\end{equation}
By the inequations (\refeq{est 1.1}) and (\refeq{est 1.2}), by Lemma
\ref{lem bound price}, and by applying Burkholder-Gundy-Davis
inequality to $\sup\limits_{0\leq t\leq T}|W^a(t)|$ and
$\sup\limits_{0\leq t\leq T}|W^b(t)|$, the claim in this lemma can
be justified. \hfill $\Box$

\begin{thm}\label{lem bound 2.1}
There exists a constant $c_4>0$, such that for any switching control
${\alpha\in\mathscr{A}}$ and any hidden order strategy
$h\in\mathscr{H}$, we have
\begin{equation}\label{goal2.5}
\begin{split}
&\expect\left[\left|\xi\left(C(T),I(T)\right)-r^C C_0\right|
\right]\\ \leq & c_4
\left((Q^a(0))^4+(Q^b(0))^4+(I_0)^2+(P^a(0))^4+(P^b(0))^4+T^4+1\right)
< \infty.
\end{split}
\end{equation}
Furthermore, for all $t\in [0,T]$,
 the value process $V(t)$ defined in (\refeq{goal2.2}) has the growth rate
\begin{equation}\label{val bound}
\begin{split}
%&-|r^I|\cdot r^F((I(t))^2+(P^a(t))^2+(P^b(t))^2+1) \leq
|V(t)|\leq c_4
\left((Q^a(t))^4+(Q^b(t))^4+(I(t))^2+(P^a(t))^4+(P^b(t))^4+(T-t)^4+1\right).
\end{split}
\end{equation}
\end{thm}
\textbf{Proof.} By Assumption \ref{assump rwd} (1), equation
(\refeq{rwd 1.3}) and equation (\refeq{invent cash 2.1}), we know
that
\begin{equation}\label{est 1.3}
|\xi(C(T),I(T))-r^C C_0|\leq |r^C|\cdot |C^h(T)+C^\alpha(T)| +
|r^I|\cdot r^F ( |z+I^h(T)+I^\alpha(T)|^2+1).
 \end{equation}
Substituting the inequalities (\refeq{est 1.4}) and (\refeq{est
1.5}) into inequality (\refeq{est 1.3}) gives the inequality
(\refeq{goal2.5}).
%If the trader never acts over the time interval
%$[t,T]$, the reward that he could get is $F(z,P^a(T),P^b(T))$. This
%reward is no larger than the supremum of the expected reward over
%all admissible trading strategies in $\mathscr{A}_t$ and
%$\mathscr{H}_t$, which is $V(t)$ by equation (\refeq{goal2.2}). By
%Assumption \ref{assump rwd}, the inequality $-r^F (z^2+1) \leq
%F(z,P^a(T),P^b(T))$ holds, hence the first inequality in (\refeq{val
%bound}) follows.
Because of the Markov property of the order book dynamics, the
inequality in (\refeq{val bound}) can be derived in the same way as
(\refeq{goal2.5}). \hfill $\Box$\\
\\ Theorem \ref{lem bound 2.1} implies that the value process $V$ defined in
(\refeq{goal2.2}) indeed exists and is finite.

\section{Relation between trading and price
switching}\label{sec equiv}

This section will state in Theorem \ref{thm equiv} the relation
between Problem \ref{prob sing} and Problem \ref{prob switch}.
Deriving from Theorem \ref{thm equiv}, the result in Proposition
\ref{prop bound} tells that the two price switching algorithms
provide lower and upper bounds of value functions of the two mixed
trading algorithms. Especially, when the premium $\epsilon$ equals
zero, the internalizing trader's optimal mixed strategy can be
achieved among the set of price switching strategies.

\begin{rmk}
The results in this section will have three implications.

(1) They help prove well-posedness of all the control problems, as
discussed at the beginning of Subsection \ref{subsec 4.2}.

(2) When the upper and lower bounds in Proposition \ref{prop bound}
do not differ much, e.g. in Fig. 6.4, the implementable lower bound
switching strategy is nearly optimal for the optimal trading
problem. Corollary \ref{cor 6.1} will show that the latter is much
harder to compute.

(3) The MiFID framework defines different types of traders and
strategies. Theorem \ref{thm equiv} compares their best expected
profits.
\end{rmk}

\begin{defn}\label{adm str 2.2} (step trading strategy)
The admissible set of step trading strategy of an internalizing
trader and that of a regular trader are denoted respectively as
$\mathscr{S}^{int}$ and $\mathscr{S}^{reg}$. Let the letter $j$
represent either ``int" or ``reg". Let
$\alpha=\{(S_n,u_n^a,u_n^b)\}_{n=0}^\infty \in \mathscr{A}^j$ be an
arbitrary admissible switching control. The processes $Z_{\alpha}^a$
and $Z_{\alpha}^b$, being the total shares that the trader has
bought and sold according to the switching control $\alpha$, are
computed from
\begin{equation}\label{str1.2.1}
Z_{\alpha}^i(t) = \sum\limits_{n: S_{n}\leq t} g^i\left(S_n,
Q^i(S_{n}-),u^i_n\right)\ind{u^i_n\geq 0}\text{, }0\leq t\leq
T\text{, }i=a,b,
\end{equation}
where the mappings $g^a$ and $g^b$ are defined in
(\refeq{rwd2.2.1}). When limit orders arrive within the spread, the
proportion of shares that the trader fills at the old price
$P^i(S_n-)$ is computed from
\begin{equation}\label{str1.2.2}
\beta_{\alpha}^i(t) =\left\{ \begin{aligned}
&\{u^i_n\}\ind{u^i_n\in[-1,0)}\text{, if }t=S_n\text{ and }N^i(S_n)-N^i(S_n-)=1;\\
& 0\text{, otherwise; }0\leq t\leq T\text{, }i=a,b.
 \end{aligned} \right.
\end{equation}
The set $\mathscr{S}^j$ of admissible step trading strategies is
defined as the collection of all the trading strategies
$Z_{\alpha}=(Z_{\alpha}^a,Z_{\alpha}^b,\beta_{\alpha}^a,\beta_{\alpha}^b)$
satisfying (\refeq{str1.2.1}) and (\refeq{str1.2.2}) for some
switching control $\alpha \in\mathscr{A}^j$. Namely,
%\begin{equation}
$\mathscr{S}^{int}= \left\{Z_{\alpha}| \alpha \in
\mathscr{A}^{int}\right\}$ and
%\end{equation}
$\mathscr{S}^{reg}= \left\{Z_{\alpha}| \alpha \in
\mathscr{A}^{reg}\right\}$.
\end{defn}
Seen from Definition \ref{adm str 2.1} and Definition \ref{adm str
2.2}, each step trading strategy of an internalizing or regular
trader is his price switching strategy denoted in terms of the total
numbers of shares bought and sold, so they are the same active
trading strategy under different names. The two definitions further
imply that
\begin{equation}
\mathscr{S}^{reg}=
\left\{Z_{\alpha}=(Z_{\alpha}^a,Z_{\alpha}^b,\beta_{\alpha}^a,\beta_{\alpha}^b)\in
\mathscr{S}^{int}| \beta_{\alpha}^a(t) \equiv \beta_{\alpha}^b(t)
\equiv 0\text{, for all }t\in[0,T]\right\}.
\end{equation}

\begin{notn}
(1) An internalizing trader's best expected reward over step trading
strategies is denoted as
\begin{equation}\label{}
V^{stp,int}(\epsilon):=\sup\limits_{Z_{\alpha}\in\mathscr{S}^{int},\,h\in\mathscr{H}}
\expect\left[\xi\left(I^{h,Z_{\alpha}}(T),C^{h,Z_{\alpha}}(T)\right)\right].
\end{equation}
(2) A regular trader's best expected reward over step trading
strategies is denoted as
\begin{equation}\label{}
V^{stp,reg}:=
\sup\limits_{Z_{\alpha}\in\mathscr{S}^{reg},\,h\in\mathscr{H}}
\expect\left[\xi\left(I^{h,Z_{\alpha}}(T),C^{h,Z_{\alpha}}(T)\right)\right].
\end{equation}
\end{notn}

\begin{thm}\label{thm equiv} The value functions of the optimal trading problem and the
 optimal switching problem have the relations\\

 (1) $V^{swt,reg}=V^{stp,reg}$;\quad\quad\quad\quad\quad\quad\quad\quad\quad
 (2) $V^{swt,int}(\epsilon)=V^{stp,int}(\epsilon)$, for all $\epsilon
\geq 0$;

 (3) $V^{stp,reg}\leq V^{mix,reg}$;
 \quad\quad\quad\quad\quad\quad\quad\quad(4) $V^{stp,int}(\epsilon)\leq V^{mix,int}(\epsilon)$, for all
$\epsilon \geq 0$;

 (5) $V^{mix,reg}\leq V^{mix,int}(\epsilon)$, for
all $\epsilon \geq 0$;
 \quad(6) $V^{swt,reg}\leq V^{swt,int}(\epsilon)$,
for all $\epsilon \geq 0$;

 (7) $V^{mix,int}(0)=V^{stp,int}(0)$;

 (8) Viewing $\epsilon$ as the
variable, the two functions $V^{swt,int}(\epsilon)$ and
$V^{mix,int}(\epsilon)$ are decreasing in $\epsilon$.
\end{thm}
Theorem \ref{thm equiv} can be proved by the results from the next
subsections \ref{subsec 5.2} and \ref{subsec 5.3}. An outline of the
proof is provided here.\\
\\\textbf{Outline of Proof of Theorem \ref{thm equiv}}
(1) and (2)  By their definitions, step trading strategies
(Definition \ref{adm str 2.2}) and price switching strategies
(Definition \ref{adm str 2.1}) have one-to-one correspondence
between each other, because the two sets in fact consist of the same
active trading strategies denoted in different terms. A price
switching strategy denotes the times of transactions and the numbers
of limits to buy and sell at each time; a step strategy denotes the
total numbers of shares bought and sold up-to-date. This explains
the first and last identities in Theorem \ref{thm equiv}.\\
(3), (4), (5) and (6) The three inequalities come from the
inclusions in Lemma \ref{lem str}.\\
(7) This identity comes from Lemma \ref{lem max 1.2}, Lemma \ref{lem
str}(1) and Proposition \ref{prop equiv 1.1}. The main idea of the
proof is to construct, in Lemma \ref{lem max 1.2}, a step trading
strategy $Z_\alpha\in\mathscr{S}^{int}$ that path-wisely replicates
the stock inventory and the cash amount produced by the mixed
trading strategy $Z\in\mathscr{Z}^{int}$. By Lemma \ref{lem str}(1),
every step trading strategy is a mixed trading strategy.
Furthermore, as will be shown by Proposition \ref{prop equiv 1.1},
an internalizing trader's two active trading strategies $Z$ and
$Z_\alpha$ path-wisely result in the same bid ask prices, replacing
the former with the latter does not change the stock inventory and
cash amount produced by a passive strategy on the hidden orders.\\
(8) The quantities $C^{h,Z}$ defined according to (\refeq{cash
3.1.1}), (\refeq{cash 3.1.2}), (\refeq{cash 3.1}) and (\refeq{invent
cash 1.1}), and $C^{h,\alpha}$ defined according to
(\refeq{rwd2.2.2})-(\refeq{rwd2.1}), viewed as functions in
$\epsilon$, are decreasing. The reward criterion $\xi$ defined in
(\refeq{rwd 1.3}) is increasing in the cash amount, because the
coefficient $r^C$ is positive. Hence $V^{mix,int}(\epsilon)$ defined
in (\refeq{exp rwd 1.2}) and $V^{swt,int}(\epsilon)$ defined in
(\refeq{goal2.3}) are decreasing in $\epsilon$.
 \hfill $\Box$

\begin{prop}\label{prop bound} The value functions of the price switching problem
provide lower and upper bounds for the value functions of the
optimal trading problem. (1) If the optimal trader is regular, then
\begin{equation}\label{rwd 1.6.9}
V^{swt,reg}\leq V^{mix,reg}\leq V^{swt,int}(0).
\end{equation}
(2) If the optimal trader is an internalizer, then
\begin{equation}\label{rwd 1.6.11}
V^{mix,int}(0)=V^{swt,int}(0)
\end{equation}
and
\begin{equation}\label{rwd 1.6.10}
V^{swt,int}(\epsilon)\leq V^{mix,int}(\epsilon)\leq
V^{swt,int}(0)\text{, for all }\epsilon \geq 0.
\end{equation}
\end{prop}
\textbf{Proof.} (1) The first inequality in (\refeq{rwd 1.6.9})
comes from Theorem \ref{thm equiv}(1)(3). The second inequality in
(\refeq{rwd 1.6.9}) comes from Theorem \ref{thm equiv}(2)(5)(7).\\
(2) By Theorem \ref{thm equiv}(2)(7), the identity (\refeq{rwd
1.6.11}) holds. The first inequality in (\refeq{rwd 1.6.10}) comes
from Theorem \ref{thm equiv}(2)(4). To prove the second inequality
in (\refeq{rwd 1.6.10}), by Theorem \ref{thm equiv}(8) it holds that
$V^{mix,int}(\epsilon)\leq V^{mix,int}(0)$, the right hand side of
which equals $V^{swt,int}(0)$ by (\refeq{rwd 1.6.11}). \hfill $\Box$

\subsection{Analysis of active strategies}\label{subsec 5.2}

It can be verified that the processes $Z^a_\alpha$, $Z^b_\alpha$,
$\beta_{\alpha}^a$ and $\beta_{\alpha}^b$ defined in
(\refeq{str1.2.1}) and (\refeq{str1.2.2}) satisfy Definition
\ref{adm str 1.1}, so every step trading strategy
$Z_{\alpha}=(Z_{\alpha}^a,Z_{\alpha}^b,\beta_{\alpha}^a,\beta_{\alpha}^b)\in
\mathscr{S}^j$ is a mixed trading strategy in the  admissible set
$\mathscr{Z}^j$, for $j=int, reg$. However, the contrary is not
true, because a mixed trading strategy can be continuous over some
time interval, but a step trading strategy is a pure jump process.
By Definition \ref{adm str 1.1}(2), a regular trader's admissible
set of mixed trading strategies is the subset of an internalizing
trader's mixed trading strategies that do not fill orders at the old
price at the time of order arrival within the spread. By their
definitions in (\refeq{set Dp}) and (\refeq{set Dr}) of Definition
\ref{adm str 2.1}, there is the set inclusion $D^{reg}\subsetneqq
D^{int}$. It follows that a regular trader's price switching
strategy is a proper subset of an internalizing trader's price
switching strategy.

\begin{lem}\label{lem str}
The admissible sets $\mathscr{S}^{int}$, $\mathscr{S}^{reg}$,
$\mathscr{A}^{int}$, $\mathscr{A}^{reg}$, $\mathscr{Z}^{int}$ and
$\mathscr{Z}^{reg}$ of trading strategies defined in Definition
\ref{adm str 2.1} and Definition \ref{adm str 2.2} have
the inclusion relations\\
 (1) $\mathscr{S}^{reg} \subsetneqq \mathscr{Z}^{reg}$;
 (2) $\mathscr{S}^{int} \subsetneqq
\mathscr{Z}^{int}$; (3) $\mathscr{Z}^{reg} \subsetneqq
\mathscr{Z}^{int}$; (4) $\mathscr{A}^{reg} \subsetneqq
\mathscr{A}^{int}$; .
\end{lem}
%\textbf{Proof.} We shall verify Definition \ref{adm str 1.1} for an arbitrary $Z_{\alpha}\in \mathscr{S}$.\\
%For any $i=a,b$ and $t\in [0,T]$, the process $Z^i_\alpha$ defined in (\refeq{str1.2.1}) and (\refeq{str1.2.2})
%indeed path-wisely has left limit and is right continuous at $t$, in other words
%``c\`{a}dl\`{a}g".
%Each summand in  (\refeq{str1.2.1}) and (\refeq{str1.2.2}) is non-negative, hence $Z^i_\alpha$
%is a non-negative and non-decreasing process. Because $S_n$ is an
%$\filtr$-stoping time and  $u_n^a$ and $u_n^b$ are
%$\filtr(S_n)$-measurable random variables for each $n=0,1,\cdots$,
%the process $Z^i_\alpha$ is adapted to the filtration $\filtr$.
%\hfill $\Box$
When the premium $\epsilon$ equals zero, the set of his admissible
step trading strategies performs equally well as an internalizing
trader's set of admissible mixed trading strategies, though the
former is a much smaller subset of the latter as stated in Lemma
\ref{lem str}(1). Whatever stock inventory and cash amount a mixed
trading strategy can produce at the terminal time, an internalizing
trader can always find a step trading strategy that path-wisely does
the same. Hence an internalizing trader's best expected reward can
be achieved over a smaller and simpler set of admissible trading
strategies. This is the role of Lemma \ref{lem max 1.2}.

\begin{lem}\label{lem max 1.2} Suppose the premium $\epsilon$ equals zero.
For any admissible mixed trading strategy
$Z=(Z^a,Z^b,\beta^a,\beta^b)\in \mathscr{Z}^{int}$, there exists an
admissible switching control
$\alpha=\{(S_n,u_n^a,u_n^b)\}_{n=0}^\infty \in \mathscr{A}^{int}$
such that the step trading strategy
$Z_{\alpha}=(Z_{\alpha}^a,Z_{\alpha}^b,\beta_{\alpha}^a,\beta_{\alpha}^b)\in
\mathscr{S}^{int}$ defined by (\refeq{str1.2.1}) and
(\refeq{str1.2.2}) for this $\alpha$ almost surely satisfies
\begin{equation}\label{redc 1.5}
I^Z(T)=I^{Z_{\alpha}}(T) \text{ and }C^Z(T)=C^{Z_{\alpha}}(T).
\end{equation}
\end{lem}
\textbf{Outline of Proof.} It suffices to construct a specific
$\alpha'=\{(S_n',u_n^{a'},u_n^{b'})\}_{n=0}^\infty \in
\mathscr{A}^{int}$ such that the step trading strategy
$Z_{\alpha'}=(Z_{\alpha'}^a,Z_{\alpha'}^b,\beta_{\alpha'}^a,\beta_{\alpha'}^b)$
defined in (\refeq{str1.2.1}) and (\refeq{str1.2.2}) for this
$\alpha'$ satisfies the identities in (\refeq{redc 1.5}). Since the
actual construction takes three pages, the detailed proof is omitted
from the paper. \hfill $\Box$

\subsection{Effect on an internalizing trader's hidden orders}\label{subsec 5.3}
Given an arbitrary admissible mixed trading strategy $Z$ from the
internalizing trader, let the step trading strategy $Z_{\alpha'}$,
for some $\alpha'=\{(S_n',u_n^{a'},u_n^{b'})\}_{n=0}^\infty \in
\mathscr{A}^{int}$, be the one constructed in Lemma \ref{lem max
1.2} to replicate the terminal stock inventory and cash amount. Let
$P^a_Z$ and $P^b_Z$ denote the price processes (\refeq{price 1.4})
controlled by the mixed trading strategy $Z$, and $P^a_{\alpha'}$
and $P^b_{\alpha'}$ denote the price processes (\refeq{price 2.4})
controlled by the switching control $\alpha'$. The construction is
such that the two strategies also produce the same times and amounts
of price change, meaning that
\begin{equation}\label{equiv price}
P^a_Z(t)=P^a_{\alpha'}(t)\text{ and
}P^b_Z(t)=P^b_{\alpha'}(t)\text{, for all }(t,\omega) \in
[0,T]\times \Omega.
\end{equation}
Let us recall that the intensities of the liquidity event processes
$H^a$ and $H^b$ are functions of the spread only. The equations
(\refeq{terminal hid}) and (\refeq{equiv price}) further imply that,
both the inventory $I^h$ and cash amount $C^h$ from an arbitrary
hidden order strategy $h=(h^a,h^b)\in\mathscr{H}$ remain the same
regardless of whether the mixed trading strategy $Z$ or the step
trading strategy $Z_{\alpha'}$ is used. The analysis in this
paragraph has verified a reinforcement of Lemma \ref{lem max 1.2},
stated as the proposition below.

\begin{prop}\label{prop equiv 1.1}
Suppose the premium $\epsilon$ equals zero. Let
$h=(h^a,h^b)\in\mathscr{H}$ be an arbitrary hidden order strategy.
For any admissible mixed trading strategy
$Z=(Z^a,Z^b,\beta^a,\beta^b)\in \mathscr{Z}^{int}$, there exists an
admissible switching control
$\alpha=\{(S_n,u_n^a,u_n^b)\}_{n=0}^\infty \in \mathscr{A}^{int}$
such that the step trading strategy
$Z_{\alpha}=(Z_{\alpha}^a,Z_{\alpha}^b,\beta_{\alpha}^a,\beta_{\alpha}^b)\in
\mathscr{S}^{int}$ defined in (\refeq{str1.2.1}) and
(\refeq{str1.2.2}) for this $\alpha$ almost surely satisfies
\begin{equation}\label{redc 1.6}
I^{h,Z}(T)=I^{h,Z_\alpha}(T) \text{ and
}C^{h,Z}(T)=C^{h,Z_{\alpha}}(T),
\end{equation}
where $I^{h,Z}(T)$, $I^{h,Z_\alpha}(T)$, $C^{h,Z}(T)$ and
$C^{h,Z_{\alpha}}(T)$ are defined in (\refeq{invent cash 1.1}) and
(\refeq{invent cash 2.1}).
\end{prop}

\section{Solving the optimal switching problem}\label{sec sol}
This section will provide the characterization of an optimal trading
strategy and derive a trading algorithm for the optimal switching
Problem \ref{prob switch}. The solution is valid regardless of
whether the trader is internalizing or regular, hence the admissible
set of switching controls is generically denoted as $\mathscr{A}$.
Before getting down to the
solution, a few notations are introduced.\\
\\The two-dimensional quantities representing both sides of the order book are
 denoted as $Q(t)=(Q^a(t),Q^b(t))$, $P(t)=(P^a(t), P^b(t))$,
$q=(q^a,q^b)$, $p=(p^a,p^b)$, $u=(u^a,u^b)$ and $h=(h^a,h^b)$ for
short. As will be shown in Theorem \ref{thm DPP}, the decision
making would only need to observe the state processes
$\{(N^a,N^b)\}_{0\leq t\leq T}$ and $\{(Q(t-),I^\alpha(t-)+I^h(t),
P(t-))\}_{0\leq t\leq T}$, which generate a smaller filtration than
$\filtr(t)$. The domain of the process
$\{(Q(t-),I^\alpha(t-)+I^h(t), P(t-))\}_{0\leq t\leq T}$ is denoted
as
\begin{equation}
\mathscr{D}= [0,\infty)^2\times\re \times \{(p^a,p^b)\in
(P^a(0),P^b(0))+\mathbb{N}^2|p^a>p^b\}.
\end{equation}
To express the change in the order book and in the inventory from
the trader's transaction, the mapping $\gamma: \Omega\times
[0,T]\times\mathscr{D}\times\mathbb{N}^2\rightarrow [0,\infty)^2
\times \re$ is defined as
\begin{equation}
\begin{split}
\gamma(t,q,z,u) =  \left( \begin{array}{l}
\ind{u^a\neq 0}\Delta^a +\ind{u^a= 0}(1-\{u^a\})q^a  \\
\ind{u^b\neq 0}\Delta^b +\ind{u^b= 0}(1-\{u^b\})q^b  \\
 z + g_\alpha(t,q,u)  \\ \end{array} \right)^\text{transpose}.
\end{split}
\end{equation}
Immediately after applying the switching control $u$ at time $t$,
the volumes and inventory become
\begin{equation}\label{rwd2.1}
\begin{split}
 (Q(t),I(t))
= \gamma\left(t,Q(t-),I^\alpha(t-)+I^h(t),u\right).
\end{split}
\end{equation}
The process $\{\int_0^t r(P(s-),h(s-)) ds\}_{0\leq t\leq T}$ defined
as
\begin{equation}
\begin{split}
r(P(t),h(t)) = & -\Delta^a h^a(t)\left(
\left(P^a(t)+P^b(t)\right)/2 \right)\lambda^a(P^a(t)-P^b(t))\\
&+ \Delta^b h^b(t) \left(\left(P^a(t)+P^b(t)\right)/2\right)
\lambda^b(P^a(t)-P^b(t))\text{, }0\leq t\leq T,
\end{split}
\end{equation} is the finite variation part in the Doob-Meyer
decomposition of the semimartingale $C^h$ defined in
$(\refeq{terminal hid})$. By the bound in (\refeq{est 1.2}), the
local martingale part of $C^h$ is a martingale. Then the value
process $V$ defined in (\refeq{goal2.2}) can be written
alternatively as
\begin{equation}\label{V rewrite}
\begin{split}
V(t)=&\sup\limits_{\alpha\in\mathscr{A}_t,\,h\in\mathscr{H}_t}\expect\bigg[r^C
\sum\limits_{n = 1}^\infty
f_\alpha\left(S_n,Q(S_{n}-),P(S_{n}-),u_{n}\right)\\
&~~~~~~~~~~~~~~~~~~~+r^C\int_t^T r(P(s-),h(s-)) ds + r^I
F(I^{h,\alpha}(T),P(T))\bigg|\filtr(t)\bigg]\text{, }0\leq t\leq T,
\end{split}
\end{equation}
because
$\expect\left[\left.\xi(I(T),C(T))\right|\filtr(t)\right]-r^C C(t)$
equals the expectation on the right hand side of the above equation.
For every $h\in\mathscr{H}_{0,t}$, the process $Y(\cdot\,;h)$ is
defined as
\begin{equation}
Y(t;h)=\int_0^t r(P(s-),h(s-)) ds + V(t) \text{, }0\leq t\leq T.
\end{equation}
For every measurable function
$\phi:[0,T]\times\mathscr{D}\rightarrow \re$, the operator
$\mathscr{M}$ is defined as
\begin{equation}\label{max 1.1}
\begin{split}
&\mathscr{M}\phi (t,q,z,p) =
\max\limits_{u\in\mathscr{U}(t,p)}\left\{
f_\alpha(t,q,p,u)+\phi\left(t,\gamma(t,q,z,u),p^a+[u^a],p^b-[u^b]\right)\right\}.
\end{split}
\end{equation}

\subsection{Optimal trading strategy}

This subsection will eventually derive, in Proposition \ref{prop str
6.1}, expressions of an optimal price switching strategy in terms of
the value process. The methodology is based on the principle that
the value process of a control problem is a supermartingale, and
becomes a martingale if and only if the control is optimal. It is
called the ``martingale method", first introduced for optimal
stopping problems in Snell (1952) and for stochastic control
problems in Davis (1979). The pivot of the arguments is the dynamic
programming principle formulated in our setting as Theorem \ref{thm
DPP}. A reference of the dynamic programming principle is Fleming
and Soner (1993). Lemma \ref{lem rc 6.1} provides the right
continuity of the value process, so that it is a qualified candidate
for using the Snell envelop technique to sequentially determine each
optimal time of trading. Lemma \ref{lem char 6.2} is the
characterization of the optimal trading strategy from the
martingality of the value process. Because Theorem \ref{lem bound
2.1} has shown that the value process is finite,  the expressions in
Proposition \ref{prop str 6.1} imply the existence of an optimal
trading strategy.

\begin{thm}\label{thm DPP} (dynamic programming principle) Given
$(Q(t-), I^\alpha(t-)+I^h(t), P(t-) )=(q,z,p)$, there exist
deterministic measurable functions $v^0$, $v^a$ and
$v^b:[0,T]\times\mathscr{D}\rightarrow \re$, and a mapping
$v:\Omega\times [0,T]\times\mathscr{D}\rightarrow \re$, such that
the value process $V$ defined by the equation (\refeq{goal2.2})
satisfies
\begin{equation}\label{DPP val 6.1}
V(t)=v(t,q,z,p) =\left\{
\begin{aligned} & v^0(t,q,z,p)\text{, if }
N^i(t)-N^i(t-)=0\text{, }i=a\text{ and }b;\\
& v^i(t,q,z,p)\text{, if } N^i(t)-N^i(t-)=1\text{, }i=a\text{ or }b.
\end{aligned} \right.
\end{equation}
The value functions $v^0$, $v^a$ and $v^b$ can be computed via the
dynamic programming principle
\begin{equation}\label{DPP 2.1.1}
\begin{split}
&v^0\left(t,Q(t-),I^\alpha(t-)+I^h(t),P(t-)\right)\\
= &\sup\limits_{(S_1, u_1) \in
\mathscr{A}_{t,1},\,h\in\mathscr{H}_{t,S_1}}\expect\bigg[ r^C
\left(f_\alpha\left(S_1,Q(S_1-), P(S_1-), u_1\right)+\int_t^{S_1} r(P(t),h(s-)) ds\right)\\
&+v\left(S_1,\gamma\left(S_1,Q(S_1-),I^\alpha(S_1-)+I^h(S_1),u_1\right),
P^a(S_1-)+[u^a_1],P^b(S_1-)-[u^b_1]\right)\bigg|\filtr(t)\bigg],
\end{split}
\end{equation}
when $N^a(t)-N^a(t-)=0$ and $N^b(t)-N^b(t-)=0$, %Especially, if the
%trader has not yet acted since the last time of price change, the
%time $T_n$ defined in (\refeq{time change}) is a time of switching,
and
\begin{equation}\label{DPP 2.1.2}
v^i\left(t,Q(t-),I^\alpha(t-)+I^h(t),P(t-)\right)=\mathscr{M}
v^0\left(t,Q(t-),I^\alpha(t-)+I^h(t),P(t-)\right)
\end{equation}
when $N^i(t)-N^i(t-)=1$, $i=a, b$. Especially, at the terminal time
$T$, the value function satisfies the terminal condition
\begin{equation}\label{DPP 2.1.3}
v\left(T,Q(T-),I^\alpha(T-)+I^h(T),P(T-)\right)=\mathscr{M}
F(I(T-),P(T-)).
\end{equation}
\end{thm}
\textbf{Proof.} The existence of the functions $v^0$, $v^a$ and
$v^b$ comes from the Markovian structure of the processes $(Q, I, P
)$, and the memoryless property of the exponential inter-arrival
times for the orders within the spread. In our context, the proof of
the dynamic programming principle is routine. To wit, take arbitrary
$\alpha\in\mathscr{A}_t$ and $h\in\mathscr{H}_t$ as defined in
Definitions \ref{adm str 2.1} and \ref{adm str 1.2}, and denote for
short
\begin{equation}\label{}
\begin{split}
\Sigma^1= & r^C \left(f_\alpha\left(S_1,Q(S_1-), P(S_1-), u_1\right)+\int_t^{S_1} r(P(t),h(s-)) ds\right);\\
\Sigma^2= & r^C\sum\limits_{n = 2}^\infty
\left(f_\alpha\left(S_n,Q(S_{n}-), P(S_{n}-),
u_{n}\right)+\int_{S_{n-1}}^{S_n} r\left(P(S_{n-1}),h(s-)\right) ds
\right)+ r^I F(I(T),P(T));\\
\Sigma^3 = &
v\left(S_1,\gamma\left(S_1,Q(S_1-),I^\alpha(S_1-)+I^h(S_1),u_1\right),
P^a(S_1-)+[u^a_1],P^b(S_1-)-[u^b_1]\right).
\end{split}
\end{equation}
Then by the same reasoning that derives equation (\refeq{V rewrite})
and by the law of iterated expectations, we have
\begin{equation}\label{ineq dpp 6.1}
\begin{split}
\expect\left[\left.\xi(I(T),C(T))\right|\filtr(t)\right]-r^C C(t)
 =  \expect\left[\left.
\Sigma^1 + \expect \left[ \Sigma^2 \left| \filtr (S_1)\right.
\right] \right|\filtr(t)\right].
\end{split}
\end{equation}
Because
\begin{equation}\label{ineq dpp 6.6}
\Sigma^3 =
\sup\limits_{\alpha\in\mathscr{A}_{S_1},\,h\in\mathscr{H}_{S_1}}
\expect \left[ \Sigma^2 \left| \filtr (S_1)\right. \right]
\end{equation}
by equations (\refeq{V rewrite}) and (\refeq{DPP val 6.1}), we know
that
\begin{equation}
\expect \left[ \Sigma^2 \left| \filtr (S_1)\right. \right] \leq
\Sigma^3.
\end{equation}
Taking supremum over $(S_1, u_1) \in \mathscr{A}_{t,1}$ on both
sides of the inequality
\begin{equation}\label{ineq dpp 6.2}
\expect\left[\left. \Sigma^1 + \expect \left[ \Sigma^2 \left| \filtr
(S_1)\right. \right] \right|\filtr(t)\right] \leq
\expect\left[\left. \Sigma^1 + \Sigma^3\right|\filtr(t)\right]
\end{equation}
and using the equations (\refeq{goal2.2}) and (\refeq{ineq dpp
6.1}), we prove that $V(t)$ is less than or equal to the right hand
side of (\refeq{DPP 2.1.1}).\\
We know from equations (\refeq{goal2.2}) and (\refeq{ineq dpp 6.1})
that
\begin{equation}\label{ineq dpp 6.3}
V(t) \geq \expect\left[\left. \Sigma^1 + \expect \left[ \Sigma^2
\left| \filtr (S_1)\right. \right] \right|\filtr(t)\right],
\end{equation}
for arbitrary $\alpha\in\mathscr{A}_t$ and $h\in\mathscr{H}_t$. The
expressions (\refeq{ineq dpp 6.6}) and (\refeq{ineq dpp 6.3}) imply
that
\begin{equation}\label{ineq dpp 6.4}
V(t) \geq \expect\left[\left. \Sigma^1
+\Sigma^3\right|\filtr(t)\right],
\end{equation}
and thus $V(t)$ greater than or equal to the right hand side of
(\refeq{DPP 2.1.1}). Both sides of the inequality hold, hence
\begin{equation}\label{ineq dpp 6.5}
\begin{split}
&V(t) = \sup\limits_{(S_1, u_1) \in
\mathscr{A}_{t,1},\,h\in\mathscr{H}_{t,S_1}}\expect\bigg[ r^C
\left(f_\alpha\left(S_1,Q(S_1-), P(S_1-), u_1\right)+\int_t^{S_1} r(P(t),h(s-)) ds\right)\\
&+v\left(S_1,\gamma\left(S_1,Q(S_1-),I^\alpha(S_1-)+I^h(S_1),u_1\right),
P^a(S_1-)+[u^a_1],P^b(S_1-)-[u^b_1]\right)\bigg|\filtr(t)\bigg].
\end{split}
\end{equation}
When $N^a(t)-N^a(t-)=0$ and $N^b(t)-N^b(t-)=0$, by (\refeq{DPP val
6.1}) there is
\begin{equation}
V(t)=v^0\left(t,Q(t-),I^\alpha(t-)+I^h(t),P(t-)\right),
\end{equation}
hence (\refeq{ineq dpp 6.5}) takes the form (\refeq{DPP 2.1.1}).
When $N^a(t)-N^a(t-)=1$ or $N^b(t)-N^b(t-)=1$ or $t=T$, the trader
has to ``trade" at time $t$, though possibly zero share, hence
(\refeq{ineq dpp 6.5}) takes the form (\refeq{DPP 2.1.2}) or
(\refeq{DPP 2.1.3}).
 \hfill $\Box$

\begin{lem}\label{lem rc 6.1}
%For a.s. $\omega\in \Omega$, and some
For every $0\leq t < t+\Delta t \leq T$, suppose the trader does not
trade over the time interval $[t, t+\Delta t]$. Then
 the processes $\left\{v^i\left(t,Q(t -),I^\alpha(t -)+I^h(t),P(t
-)\right)\right\}_{0\leq t\leq T}$, $i=0,a,b$, are continuous in the
time $t$, meaning that
\begin{equation}\label{eqn rc}
\begin{split}
 &\lim\limits_{|t-t'|\rightarrow 0+} \left|v^i\left(t,Q(t -),I^\alpha(t -)+I^h(t),P(t
-)\right) -
v^i\left(t',Q(t'-),I^\alpha(t'-)+I^h(t'),P(t'-)\right)\right|=0.
\end{split}
\end{equation}
\end{lem}
\textbf{Proof.} It suffices to prove the continuity of $v^0$, then
the continuity of $v^a$ and $v^b$ follows from the expression
(\refeq{DPP 2.1.2}).

Take two arbitrary times $t\leq t' \in [0,T]$, an arbitrary price
switching strategy $\alpha \in \mathscr{A}_{t}$ and an arbitrary
hidden order strategy $h\in \mathscr{H}_{t}$. Denote $\Delta t:=
t'-t$. Suppose $N^a(t)-N^a(t-)=0$ and $N^b(t)-N^b(t-)=0$.

 (continuity in the volume $q$) For any two sets of initial
values $(q,z,p)$ and $(q',z,p)\in \mathscr{D}$, the resulted state
processes are respectively denoted as $(Q,I,P)$ and $(Q',I',P')$. If
the trader never trades, then, taking the ask side for example,
there are three possibilities of the dynamics.

(1) Limit sell orders arrive in the spread before either $Q^a(\cdot)
= q^a+\sigma^a (W^a(\cdot)-W^a(t))$ or $Q^{a'}(\cdot) =
q^{a'}+\sigma^a (W^a(\cdot)-W^a(t))$ reaches zero, in which case the
new ask prices $P^a$ and $P^{a'}$ equal $p^a-1$ and the volumes
$Q^a$ and $Q^{a'}$ both become $\Delta^a$ at the time of arrival.

(2) Limit sell orders does not arrive in the spread before both
$Q^a(\cdot)$ and $Q^{a'}(\cdot)$ reach zero. At the time of arrival
the new ask prices $P^a$ and $P^{a'}$ equal $p^a+1$ and the volumes
$Q^a$ and $Q^{a'}$ still differ by $\left|q^a-q^{a'}\right|$.

(3) Limit sell orders arrive in the spread when one of $Q^a(\cdot)$
and $Q^{a'}(\cdot)$ has reached zero and the other one has not.

In cases (1) and (2), the difference in the trader's stock inventory
does not exceed $\left|q^a-q^{a'}\right|+\left|q^b-q^{b'}\right|$,
and that in the cash amount does not exceed
$\left(\bar{p}^a+\epsilon\right)\left|q^a-q^{a'}\right|
+\left(\underline{p}^b+\epsilon\right)\left|q^b-q^{b'}\right|$. The
probability that type (3) events ever happen over the entire time
horizon $[0,T]$ converges to zero, as
$\left|q^a-q^{a'}\right|\rightarrow 0$. By Assumption \ref{assump
rwd} (2) and from the bounds of the price, stock inventory and cash
amount in Lemma \ref{lem bound price} and Lemma \ref{lem bound inven
cash}, we know that
\begin{equation}\label{cont lem 6.1.3}
v^0(t,q,z,p)=\lim\limits_{q'\rightarrow q}v^0(t,q',z,p).
\end{equation}

(continuity in the time $t$) Take any initial values $(q,z,p)\in
\mathscr{D}$ and any number $\Delta t \in [0,T- t]$. Let
$\mathscr{A}_t(T-\Delta t)$ denote the active trading strategies
with the terminal time $T$ replaced by $T-\Delta t$.  Then there is
the relation
\begin{equation}\label{cont lem 6.1.6}
v^0(t+\Delta t,q,z,p)= \left( \sup\limits_{\alpha \in
\mathscr{A}_t(T-\Delta t), h \in
\mathscr{H}_t}\expect\left[\left.\xi(I(T-\Delta t),C(T-\Delta
t))\right|\filtr(t)\right]-r^C C(t) \right).
\end{equation}
Because the change in the order book dynamics during the time
interval $[T- \Delta t, T]$ is of the order $O(\Delta t)$, the best
expected reward from terminating at the time $T- \Delta t$ or at the
time $T$ differs up to $O(\Delta t)$, meaning that
\begin{equation}\label{cont lem 6.1.4}
v^0(t,q,z,p)=  v^0(t+\Delta t,q,z,p) + O(\Delta t).
\end{equation}

The continuity of the process $\left\{v^0\left(t,Q(t -),I^\alpha(t
-)+I^h(t),P(t -)\right)\right\}_{0\leq t\leq T}$ can be concluded
from the identities (\refeq{cont lem 6.1.3}) and (\refeq{cont lem
6.1.4}) and from the properties of Brownian motions and Poisson
processes that drive the state
process. \hfill $\Box$\\
\\The proofs of Lemma \ref{lem char 6.2} and Proposition \ref{prop
str 6.1} follow the routine procedure on how to characterize the
optimal control and optimal stopping time via the martingale method.
Because they are very long, the proofs are not provided here.
Interested readers could find the original idea in Davis (1979) and
Snell (1952), and the arguments for a most similar result in Section
2.2.2 of Li (2011).

\begin{lem}\label{lem char 6.2}
The price switching strategy $\alpha^*=(S_1^*,u^*_1) \in
\mathscr{A}_{t,1}$ and hidden order strategy $h^* \in
\mathscr{H}_{t,S_1^*}$ achieve the supremum in (\refeq{DPP 2.1.1}),
if
and only if all of the four conditions below hold.\\

(1) $\{Y(t\,;h)\}_{0\leq t\leq T_1}$ is a supermartingale, for every
$h \in \mathscr{H}_{t,T_1}$;

(2) $\{Y(t\wedge S_1^*\,;h^*)\}_{0\leq t\leq T_1}$ is a martingale;

(3) either
$\left(v^0-\mathscr{M}v^0\right)\left(S_1^*,Q(S_1^*-),I^{\alpha^*}(S_1^*-)+I^{h^*}(S_1^*),P(S_1^*-)\right)=0$,
or $S_1^*=T_1$;

(4) $u^{*}_1  =  \,arg
\max\limits_{u\in\mathscr{U}(S^*_1,P(t))}\bigg\{r^C
\left(f_\alpha(S^*_1,Q(S^*_1-),P(t),u^*_1) +\int_t^{S^*_1}
r(P(t),h^*(s-)) ds\right)$

$~~~~~~~~~~~~+v\left(S^*_{1},\gamma(S^*_{1},Q(S^*_{1}-),I^{\alpha^*}(S^*_{1}-)+I^{h^*}(S^*_{1}),u^*_1),
P^a(t)+[u^{a*}_1],P^b(t)-[u^{b*}_1]\right)\bigg\}.$
\end{lem}
%\textbf{Proof.} \hfill $\Box$

\begin{prop}\label{prop str 6.1} There exist an optimal switching control $\alpha^* = \{(S_n^*,u^{a*}_n,u^{b*}_n)\}_{n=1}^\infty \in \mathscr{A}$
and an optimal hidden order strategy $h^* = (h^{a*},h^{b*}) \in
\mathscr{H}$, which are defined in the following way. Let
$S^*_0=u^{a*}_0=u^{b*}_0=0$. For $n=1,2,\cdots$, the optimal trading
time $S^*_n$ can be expressed as
\begin{equation}\label{}
S^*_n =\left\{
\begin{aligned} & \inf \left\{S^*_{n-1}< t \leq T\big|
\left(v^0-\mathscr{M}v^0\right)\left(t,Q(t-),I^{\alpha^*}(t-)+I^{h^*}(t),P(t-)\right)=0%\text{, }\\
%&~~~~~~~~~~~~~~~~~~~~~~~~~ N^a(t)-N^a(t-)=1\text{, or }N^b(t)-N^b(t-)(t)=1
\right\}\wedge T_n,\\
&~~~~~\text{if
}S^*_{n-1}< T;\\
&T+1\text{, if }S^*_{n-1}=T.
\end{aligned}
\right.
\end{equation}
If $S^*_n=T+1$ , then $u^{a*}_n=u^{b*}_n=0$; otherwise
\begin{equation}\label{}
\begin{split}
& u^*_n  =  \,arg
\max\limits_{u\in\mathscr{U}(S^*_{n},P(S^*_{n-1}))}\bigg\{
r^C f_\alpha(S^*_{n},Q(S^*_{n}-),P(S^*_{n-1}),u)\\
&+v\left(S^*_{n},\gamma(S^*_{n},Q(S^*_{n}-),I^{\alpha^*}(S^*_{n}-)+I^{h^*}(S^*_{n}),u),
P^a(S^*_{n-1})+[u^a],P^b(S^*_{n-1})-[u^b]\right)\bigg\}.
\end{split}
\end{equation}
Denoting as
\begin{equation}\label{}
\begin{split}
&r^0\left(t,Q(t),I^{\alpha^*}(t)+I^{h^*}(t-),P(t),h(t)\right)\\
:=&\left(v^0\left(t,Q(t),I^{\alpha^*}(t)+I^{h^*}(t-)+\Delta^a,P(t)\right)
-v^0\left(t,Q(t),I^{\alpha^*}(t)+I^{h^*}(t-),P(t)\right)\right)\\
& \cdot h^a(t)\,\lambda^a\left(P^a(t)-P^b(t)\right)\\
+ &
\left(v^0\left(t,Q(t),I^{\alpha^*}(t)+I^{h^*}(t-)-\Delta^b,P(t)\right)
-v^0\left(t,Q(t),I^{\alpha^*}(t)+I^{h^*}(t-),P(t)\right)\right)\\
& \cdot h^b(t)\,\lambda^b\left(P^a(t)-P^b(t)\right),
\end{split}
\end{equation}
for $0\leq t\leq T$, the optimal hidden order strategy $h^*$ can be
expressed as
\begin{equation}\label{}
\begin{split}
& h^*(t)  =  \,arg \max\limits_{h(t)\in \{0,1\}^2}
\left\{r\left(P(t),h(t)\right)+r^0\left(t,Q(t),I^{\alpha^*}(t)+I^{h^*}(t-),P(t),h(t)\right)
\right\}.
\end{split}
\end{equation}

\end{prop}

%$\left(t,Q(t-),I^\alpha(t-)+I^h(t),P(t-)\right)$

\subsection{Numerical algorithm}\label{subsec numeric}

This subsection will present the numerical algorithm to compute the
value function and optimal trading strategy for the discretized
version of the optimal price switching Problem \ref{prob switch}. We
shall specify different complexities of this algorithm on a serial
computer and on a GPU cluster. \\
\\The time and the state process are discretized over a grid
$\mathscr{T}\times\mathscr{X}$, where $\mathscr{T}$ as the grid for
the time $t\in[0,T]$ is defined as
\begin{equation}
\begin{split}
\mathscr{T}=  & \{0=t_0,t_1,t_2,\cdots,t_K = T\} = \{0,\Delta t,
2\Delta t,\cdots,K\Delta t = T \}
\end{split}
\end{equation}
and $\mathscr{X}$ as the grid for the state process $(Q,I,P)$ is a
bounded set in $\mathscr{D}$ with $|\mathscr{X}|<\infty$ elements.
When the grid tends finer and finer, the limit
\begin{equation}
\lim\limits_{K\rightarrow \infty,\, |\mathscr{X}|\rightarrow
\infty}\mathscr{T}\times\mathscr{X}
\end{equation}
is assumed to be a dense set in  $[0,T]\times\mathscr{D}$.\\
\\The algorithm takes the three steps to be specified below.
Its outputs will be the value function and the optimal trading
strategy $(\bar{v}^0(t_k,x),\bar{u}^{*0}(t_k,x),\bar{h}^{*}(t_k,x))$
when limit orders do not arrive within $(t_{k-1},t_k]$, and
$(\bar{v}^i(t_k,x),\bar{u}^{*i}(t_k,x))$ when limit sell (buy)
orders arrive within $(t_{k-1},t_k]$, for all
$(t_k,x)\in\mathscr{T}\times\mathscr{X}$. The pseudo codes are provided in Table 6.1.\\
\\Step \textbf{1.} (at the terminal time) At the time $t_K=T$ , the
terminal condition is
$\bar{v}(T,x)=\mathscr{M}F(z,p)$.\\

Step \textbf{1.1} Compute the reward from trading at the terminal
time for every trading strategy $u\in\mathscr{U}(T,p)$
%and $h\in\{0,1\}^2$
as
\begin{equation}\label{alg 1.1}
\bar{v}(T,x;u)=\bar{v}^0(T,x;u)=\bar{v}^i(T,x;u)=f_\alpha(T,q,p,u)+F(z+g_\alpha(T,q,u),p).
\end{equation}

Step \textbf{1.2} The maximum reward from trading at the terminal
time is
\begin{equation}\label{alg 1.2.1}
\bar{v}(T,x)=\bar{v}^0(T,x)=\bar{v}^i(T,x)=\max\{\bar{v}(T,x;u)|u\in\mathscr{U}(T,p)\}.
\end{equation}

The optimal trading strategy is
\begin{equation}\label{alg 1.2.2}
\bar{u}^*(T,x)=\{u\in\mathscr{U}(T,p)\,|\text{ such that }
\bar{v}(T,x;u)=\bar{v}(T,x)\}.
\end{equation}
\hfill $\Box$\\
Step \textbf{2.} (simulate the controlled state process) %For every $t_k\in
%\mathscr{T}$ and $x=(q^a,q^b,z,p^a,p^b)\in \mathscr{X}$,
With the initial values
\begin{equation}\label{alg 2.1}
\begin{split}
X_{t_k,x}(t_{k};u):=\left(Q_{t_k,x}(t_{k};u),I_{t_k,x}(t_{k};u),P_{t_k,x}(t_{k};u)\right)
=  \left(\gamma(t_k,q,z,p,u),p^a+[u^a],p^b-[u^b]\right),
\end{split}
\end{equation}
simulate the state process
\begin{equation}\label{disc 1.1}
X_{t_k,x}(t_{k+1};u,h)=\left(
Q_{t_k,x}(t_{k+1};u),I_{t_k,x}(t_{k+1};u,h),
P_{t_k,x}(t_{k+1};u)\right)
\end{equation}
according to
\begin{equation}\label{disc 1.2}
\left\{
\begin{aligned}
\tilde{Q}^i_{t_k,x}(t_{k+1};u)= & \,Q^i_{t_k,x}(t_{k};u) + \sigma^i
\times \text{ (a Normal r.v. with mean zero and variance }\Delta
t\text{);}\\
Q^i_{t_k,x}(t_{k+1};u)= & \,\tilde{Q}^i_{t_k,x}(t_{k+1};u) \ind{\tilde{Q}^i_{t_k,x}(t_{k+1};u)>0}
+ \Delta^i \ind{\tilde{Q}^i_{t_k,x}(t_{k+1};u)\leq 0} ;\\
I_{t_k,x}(t_{k+1};u,h)= & \,I_{t_k,x}(t_{k};u) + \Delta^a h^a \times
\text{ (a Poisson r.v. with intensity }\lambda^a(p^a-p^b)\Delta t\text{)}\\
&~~~~~~~~~~~~~~-\Delta^b h^b \times \text{ (a Poisson r.v. with
intensity }\lambda^b(p^a-p^b)\Delta t\text{);}\\
P^a_{t_k,x}(t_{k+1};u)= & \,p^a +[u^a] + \ind{\tilde{Q}^i_{t_k,x}(t_{k+1};u)\leq 0};\\
P^b_{t_k,x}(t_{k+1};u)= & \,p^b -[u^b] -
\ind{\tilde{Q}^i_{t_k,x}(t_{k+1};u)\leq 0}.
\end{aligned}\right.
\end{equation}
So that the state process remains within the grid $\mathscr{X}$, the
truncated value from each simulation $\bar{X}_{t_k,x}(t_{k+1})$ is
obtained from
\begin{equation}\label{disc 1.3}
\bar{X}_{t_k,x}(t_{k+1};u,h):= arg\min
\big\{\left|X_{t_k,x}(t_{k+1};u,h)-y \right| \big| y\in
\mathscr{X}\big\}.
\end{equation}
(It is possible to directly simulate the truncated values.)\\
Run $M$ simulations to get $\bar{X}_{t_k,x}(t_{k+1};u,h)$ according
to the equations (\refeq{disc 1.1}), (\refeq{disc 1.2}) and
(\refeq{disc 1.3}). The $M$ simulated values  are denoted as
$\left\{\bar{X}^m_{t_k,x}(t_{k+1};u,h)\right\}_{m=1}^M$. For
$i=a,b$, simulate $M$ Poisson random variables
$\left\{N^i_m\right\}_{m=1}^M$ with the intensity
$\theta^i(p^a-p^b)\Delta t$ to represent whether limit orders arrive
within the spread during the time interval $(t_k,t_{k+1}]$. \hfill
$\Box$
\\\\Step \textbf{3.} (value function and optimal trading strategy)
This step conducts the optimization procedure by the dynamic
programming principle
\begin{equation}\label{bwd 1.1}
\begin{split}
\bar{v}(t_{k},x)= & \max\limits_{u\in \mathscr{U}(t_{k},p),\,h\in
\{0,1\}^2} \bigg\{\left( \frac{p^a+p^b}{2} h^b\lambda^b(p^a-p^b)-
\frac{p^a+p^b}{2} h^a\lambda^a(p^a-p^b)\right)\Delta
t\\
&~~~~~~~~~~~~~~~~~~~~~~~~~~~+f_\alpha(t_k,q,p,u)+\expect\left[
\bar{v}(t_{k},\bar{X}_{t_k,x}(t_{k+1};u,h)) \right]\bigg\}.
\end{split}
\end{equation}
%\hfill $\Box$\\

Step \textbf{3.1} (approximating the expectation) The conditional
expectation\\ $\expect\left[
\bar{v}(t_{k},\bar{X}_{t_k,x,}(t_{k+1})) \right]$ in (\refeq{bwd
1.1}) is approximated by computing
\begin{equation}\label{alg 2.1.1}
\begin{split}
& \hat{v}(t_{k},x;u,h)\\
:= & \frac{1}{M}\sum\limits_{m=1}^M
\left(\ind{N^a_m=0,N^b_m=0}\bar{v}^0\left(t_{k},\bar{X}^m_{t_k,x}(t_{k+1};u,h)\right)
+\sum\limits_{i=a,b}\ind{N^i_m=0}\bar{v}^i\left(t_{k},\bar{X}^m_{t_k,x}(t_{k+1};u,h)\right)\right).
\end{split}
\end{equation}

Step \textbf{3.2} (value function and trading strategy when no
arrival within the spread) This is the case when there is no limit
order arrival within the spread throughout the time interval
$(t_{k-1},t_{k}]$, meaning that $N^i(t_{k})-N^i(t_{k-1})=0$, for
$i=a$ and $b$. The reward from using a generic trading strategy
$u\in\mathscr{U}(t_k,p)$ and $h\in\{0,1\}^2$ is
\begin{equation}\label{alg 2.2.1}
\begin{split}
\bar{v}^0(t_{k},x;u,h)= & \left( \frac{p^a+p^b}{2}
h^b\lambda^b(p^a-p^b)-\frac{p^a+p^b}{2}
h^a\lambda^a(p^a-p^b)\right)\Delta t\\
&+f_\alpha(t_k,q,p,u)+ \hat{v}(t_{k},x;u,h).
\end{split}
\end{equation} The optimal value from
trading is
\begin{equation}\label{alg 2.2.2}
\bar{v}^0(t_{k},x)=
\max\left\{\bar{v}^0(t_{k},x;u,h)|u\in\mathscr{U}(t_k,p) \text{ and
} h\in\{0,1\}^2\right\}.
\end{equation}
The optimal trading strategy is
\begin{equation}\label{alg 2.2.3}
(\bar{u}^{0*}(t_{k},x),\bar{h}^*(t_{k},x))= \left\{\left.
u\in\mathscr{U}(t_k,p) \text{ and } h\in\{0,1\}^2\right| \text{ such
that } \bar{v}^0(t_{k},x;u,h)=\bar{v}^0(t_{k},x)\right\}.
\end{equation}

%\hfill $\Box$\\

Step \textbf{3.3} (value function and trading strategy when there is
arrival within the spread) This is the case when limit orders arrive
within the spread at some point during the time interval
$(t_{k-1},t_{k}]$, meaning that $N^i(t_{k})-N^i(t_{k-1})=1$, for
$i=a$ or $b$. The reward from using a generic trading strategy
$u\in\mathscr{U}(t_k,p)$ is
\begin{equation}\label{alg 3.3.1}
\bar{v}^i(t_{k},x;u)= f_\alpha(t_k,q,p,u)+ \bar{v}^0(t_{k},x;u)
\end{equation}
The optimal value from trading is
\begin{equation}\label{alg 3.3.2}
\bar{v}^i(t_{k},x)=
\max\left\{\bar{v}^i(t_{k},x;u)|u\in\mathscr{U}(t_k,p)\right\}.
\end{equation}
The optimal trading strategy is
\begin{equation}\label{alg 3.3.3}
\bar{u}^{i*}(t_{k},x)= \left\{u\in\mathscr{U}(t_k,p)\,| \text{ such
that } \bar{v}^i(t_{k},x;u)=\bar{v}^i(t_{k},x)\right\}.
\end{equation}
\hfill $\Box$\\
\\We would like to distinguish between the computational
complexities of the backward induction algorithm on a CPU and on a
GPU. The corollary below draws conclusion from the discretization
via the dynamic programming of the price switching problem, which is
of the type of combined impulse control and optimal control.
Interested readers may verify if the result is the same for other
methods (PDE or backward SDE, if applicable) and other control
types.

\begin{cor} \label{cor 6.1} Let $|\mathscr{T}|$, $|\mathscr{X}|$ and $|\mathscr{U}|$
respectively be the mesh sizes of the discretized time grid, space
grid and admissible control set, and $M$ be the number of simulation
paths to estimate the conditional expectation. Using serial
computation, the time complexity of the algorithm is $|\mathscr{T}|
\times |\mathscr{X}| \times |\mathscr{U}|\times M$. Using parallel
computation to switch as much as possible the complexity to space
complexity, the space complexity of the algorithm is $|\mathscr{X}|
\times |\mathscr{U}|\times M$, and the time complexity can be
reduced to at most $|\mathscr{T}| \times |\mathscr{U}|$ .
\end{cor}
\textbf{Proof.} Table 6.1 lists the pseudo codes of the algorithm,
the computational complexity at every step and whether it is
parallelizable or not. The computation at every node in the state
space $\mathscr{X}$ is always parallelizable, because it uses
results from the previously computed time step, and does not use any
other nodes at the same time step. The number $|\mathscr{U}|$ is the
complexity to get the maximum expected reward among all admissible
controls, hence it cannot be carried out in parallel. For example,
to get the maximum among the numbers $\{a_1,a_2,\cdots,a_N\}$, one
can inductively compute $b_1:=a_1$ and $b_n:=\max\{b_{n-1},a_n\}$,
for $n=2,\cdots,N$. Then $b_N=\max\{a_1,a_2,\cdots,a_N\}$. \hfill
$\Box$

\begin{rmk}
There are two important observations from Corollary \ref{cor 6.1}.

(1) Because all the nodes in the state space at every time step can
be computed in parallel, stochastic control problems of mediumly
high dimension is no longer numerically forbidding.

(2) The minimum time complexity on a GPU cluster is $|\mathscr{T}|
\times |\mathscr{U}|$, which is the number of time steps multiplies
the size of the admissible control set. The admissible set of mixed
strategies is the cube $\left[0,\bar{p}^a\right]\times
\left[0,\underline{p}^b\right]$, while the admissible set of price
switching strategies is only the integer points inside that cube,
hence the price switching problem is much simpler to implement.
\end{rmk}

\subsection{Implementation}
This subsection implements the algorithm in a simpler Binomial
model, to the best capability of the author's PC. An interesting
application of the numerical results would be to calculate a ``fair"
internalization premium $\epsilon^*$. \\
\\From every time step $t_k$ to $t_{k+1}$, the randomness in the model
is captured by six Binomial variables. Independence is assumed
unless mentioned otherwise. Other features remaining the same, the
modifications from the previous subsection are the following.

(1) The change in the volume $Q^i$ caused by market participants
other than the trader is a random variable $R(Q^i)\in \{-1,1\}$ with
probabilities $\{0.5,0.5\}$, $i=a,b$.

(2) Let the pair of random variables $(R(N^a),R(N^b))\in
\{(0,0),(1,0),(0,1)\}$ indicate whether limit sell and buy orders
arrive (value one) or not (value zero) at one tick below the ask
price and one tick above the bid price, when the spread is greater
than one tick. The three scenarios are assigned probabilities
$\{1-p_N, p_N/2, p_N/2\}$, where $p_N = 0.3 \cdot
\min\{\text{spread}-1, 1\}$.

(3) Let the pair of random variables $(R(H^a),R(H^b))\in
\{(0,0),(1,0),(0,1)\}$ indicate whether there is a liquidity event
(value one) or not (value zero) that consumes the trader's hidden
buy and sell orders at the mid price. The three scenarios are
assigned probabilities $\{0.5, 0.25, 0.25\}$.

(4) In addition, internalization here means only filling $\Delta^a$
($\Delta^b$) shares at the time $t_k$ price $P^a(t_k)$ ($P^b(t_k)$),
when at time $t_{k+1}$ limit sell (buy) orders arrive at a better
price $P^a(t_{k+1})=P^a(t_k)-1$
($P^b(t_{k+1})=P^b(t_k)+1$), indicated by $R(N^a)=1$ ($R(N^b)=1$).\\
\\We use $\Delta^a=\Delta^b=5$, $\underline{p}^b=12$ and $\bar{p}^a=18$. The time mesh is
\begin{equation}
\{t_0,t_1,\cdots,t_K\}=\{1,2,\cdots,10\}.
\end{equation}
At the terminal time $t_K=10$, the trader's stock inventory is
valued at $P^b(10)-2$ per share if it's positive, and $P^a(10)+2$
per share if negative.\\
\\Because the state space is infinite, it has to be truncated somehow
on the boundary. The largest grid that the author's PC can accept is
\begin{equation}
\left(Q^a,Q^b,I,P^a,P^b\right)\in \mathscr{X} =
\{0,1,\cdots,9,10\}^2 \times \{-20,-19,\cdots,19,20\} \times
\{12,13,\cdots,17,18\}^2.
\end{equation}
The grid contains 104181 admissible points where $P^a>P^b$.\\
\\How the trader's optimal trading strategy interacts with simulated
price paths is illustrated in Fig. 6.1 (regular trader), Fig. 6.2
(systemic internalizer) and Fig. 6.3 (systemic internalizer). The
initial time is $t_0=1$ and terminal time $T=10$. The initial values
are $Q^a(1)=Q^b(1)=5$, $P^a(1)=16$, $P^b(1)=15$ and $I(1)=0$. The
trader's activities in all three figures display an attempt to sell
short and push down the price. He indeed uses combinations of
different order types - active, hidden and internalizing orders. Due
to the truncation on the inventory, it is however not quite
informative to compute the profit. More advanced devices are needed
to allow for a
wider grid, especially a larger range of the inventory variable.\\
\\It is interesting to see the effect of internalization on the trader's
best expected profit. The relative difference in best expected
profits between a systemic internalizer and a regular trader is
defined as
\begin{equation}\label{relative diff}
\begin{split}
& V^{diff}(t_0;q^a,q^b,z,p^a,p^b;\epsilon)\\
:=&\left(V^{int}(t_0;q^a,q^b,z,p^a,p^b;\epsilon)-V^{reg}(t_0;q^a,q^b,z,p^a,p^b)\right)/V^{reg}(t_0;q^a,q^b,z,p^a,p^b),
\end{split}
\end{equation}
where $V^{int}(t_0;\cdot;\epsilon)$ and $V^{reg}(t_0;\cdot)$ are
respectively the time-$t_0$ value functions of systemic internalizer
and regular trader defined in (\refeq{goal2.2}), and $\epsilon$ is
the internalization premium. Fig. 6.4 shows the distribution of
$V^{diff}(t_0;q^a,q^b,z,p^a,p^b;\epsilon=0)$, with the initial
values $(q^a,q^b,z,p^a,p^b)$ ranging over the 104181 admissible
points on the grid. The systemic internalizer's best expected profit
is $1\%$-$15\%$ higher than that of the regular trader on about
$35\%$ of the admissible points. This means that internalization,
when applied in the right situations, can be on average profitable.
Furthermore, it suggests a way to specify a ``fair" value
$\epsilon^*$ of the internalization premium, so that both parties of
the transaction will gain.

\begin{rmk}
Let a weight function $w: \mathscr{X}\rightarrow (0,1)$;
$(q^a,q^b,z,p^a,p^b)\mapsto w(q^a,q^b,z,p^a,p^b)$ represent the
likelihood of each point in the state space, satisfying
\begin{equation}
\sum\limits_{(q^a,q^b,z,p^a,p^b)\in\mathscr{X}}w(q^a,q^b,z,p^a,p^b)
= 1.
\end{equation}

For some commonly recognized reward criterion $F$ in equation
(\refeq{rwd 1.3}), some typical duration $T$ of a trading period and
some proper grid $\mathscr{X}$ of the state space, the ``fair"
internalization premium $\epsilon^*$ should be a strictly positive
number such that the weighted average
\begin{equation}\label{opt premium}
\sum\limits_{(q^a,q^b,z,p^a,p^b)\in\mathscr{X}}V^{diff}(t_0;q^a,q^b,z,p^a,p^b;\epsilon^*)w(q^a,q^b,z,p^a,p^b)
\end{equation}
is somewhere above zero.

Since internalization is an additional choice that brings a higher
best expected profit, $\epsilon^*$ should be positive. Since
internalization provides price improvements to his counterpart,
$\epsilon^*$ should be low enough to keep it profitable for a
systemic internalizer to do so.
\end{rmk}

\section*{Acknowledgements}
I would like to thank many people from conversations with whom I
started to learn about market microstructure. I am mostly grateful
to Charles-Albert Lehalle at Cr\'{e}dit Agricole Cheuvreux and
Capital Fund Manegement, for sharing with me his insights into
market microstructure and algorithmic trading, through discussions
during the breaks at academic seminars, in his office and via
numerous emails. His insights provided a guideline of where the
mathematics should go at every step of the work. In this paper, the
title, the limit order book model, the idea of a ``fair"
internalization premium and most of the deep understandings about
the literature and about the European market should be attributed to
Charles-Albert. Besides, the course on programming GPU in CUDA/C
language, taught by Dr. Lokman A. Abbas
Turki at TU Berlin, gave me much more concrete knowledge about parallel computation. \\
\\A very primitive idea of this paper was initiated when I was graduating from Columbia
University in New York, while the actual work took place
consecutively  at Universit\'{e} d'Evry, at Humboldt-Universit\"{a}t
zu Berlin and during my current transition. I would like to
acknowledge the financial support kindly provided by l'Institut
Europlace de Finance and l'Institut Louis Bachelier Paris, by
Matheon Berlin and by my mother.

\newpage
\section*{Table}
\sffamily \setcounter{section}{6} \setcounter{table}{0}
\begin{center}
\begin{table}[h!]\label{table codes}
\caption{The parallelizable algorithm in pseudo codes}
\begin{tabular}{|lll|}
\hline\\
Pseudo Codes    &Computational Complexity& Parallelizable?\\
\hline
&&\\
for ($x$ in $\mathscr{X}$)  &&\\
\{   &&\\

\hspace{6pt}for ($u$ in $\mathscr{U}(T,p)$)&&\\

\hspace{3pt}\{ &&\\

\hspace{9pt}Step 1.1 & $|\mathscr{X}|\times|\mathscr{U}|$ & yes\\

\hspace{3pt}\} &$+$&\\
Step 1.2 & $|\mathscr{X}|\times|\mathscr{U}|$ & in $x$\\
print $\bar{v}(T,x)$ and $u^{*}(T,x)$ to file &&\\
\} &$+$&\\
&&\\for($k=K-1$, $k--$, $k\geq 0$) &$|\mathscr{T}|$&no\\
\{ &$\times$&\\

\hspace{6pt}for ($x$ in $\mathscr{X}$)&$\frown$&\\

\hspace{3pt}\{&&\\

\hspace{9pt}for ($u$ in $\mathscr{U}(t_k,p)$ and $h$ in
$\{0,1\}^2$)&$|\mathscr{X}|\times|\mathscr{U}|\times M$&yes\\

\hspace{12pt}\{&&\\

\hspace{15pt}Step 2&&\\

\hspace{15pt}Step 3.1& $+$ &\\

\hspace{15pt}equation (\refeq{alg 2.2.1}) in Step 3.2&&\\

\hspace{12pt}\}&&\\

\hspace{9pt}equation (\refeq{alg 2.2.2}) in Step 3.2&$|\mathscr{X}|\times|\mathscr{U}|$  & in $x$\\
&$+$&\\
\hspace{9pt}equation (\refeq{alg 2.2.3}) in Step 3.2&$|\mathscr{X}|\times|\mathscr{U}|$ & in $x$\\

\hspace{9pt}print $\bar{v}^0(t_{k},x)$, $u^{0*}(t_{k},x)$ and
$h^*(t_{k},x)$ to file&&\\

\hspace{3pt} \}&&\\
&&\\
\hspace{6pt}for ($x$ in $\mathscr{X}$)&$+$&\\

\hspace{3pt}\{&&\\

\hspace{9pt}for ($u$ in $\mathscr{U}(t_k,p)$)&&\\

\hspace{9pt}\{&&\\

\hspace{15pt}equation (\refeq{alg 3.3.1}) in Step 3.3&$|\mathscr{X}|\times|\mathscr{U}|$&yes\\

\hspace{9pt}\}&$+$&\\

\hspace{9pt}equation (\refeq{alg 3.3.2}) in Step 3.3&$|\mathscr{X}|\times|\mathscr{U}|$  & in $x$\\
&$+$&\\
\hspace{9pt}equation (\refeq{alg 3.3.3}) in Step 3.3&$|\mathscr{X}|\times|\mathscr{U}|$ & in $x$\\

\hspace{9pt}print $\bar{v}^i(t_{k},x)$ and $u^{i*}(t_{k},x)$ to
file&$\smile$&\\
\hspace{3pt}\}&&\\
\}&&\\&&\\
\hline
\end{tabular}
\end{table}
\end{center}
\normalfont

\newpage
\section*{Figures}\label{subsec Append 3}

\subsection*{Figures from Section \ref{sec dynamics}}
\setcounter{section}{2} \setcounter{figure}{0}
\begin{center}
\begin{figure}[h!]\label{order book}
\setlength{\unitlength}{0.5cm}
\begin{picture}(22,17)
%% coordinate and labels
\put(8,14.5){\text{Light Market}} \put(21,14.5){\text{Dark Pool}}
\put(5,1){\vector(0,1){14}} \put(1.6,14){\text{Price}}
\put(5,2){\vector(1,0){18}} \put(20,1){\text{Volume}}
\put(1.6,3.8){$P^b(t)-2\delta$} \put(1.6,4.8){$P^b(t)-\delta$}
\put(1.6,5.8){$P^b(t)$} \put(1.6,6.8){$P^b(t)+\delta$}
\put(1.6,11.8){$P^a(t)+2\delta$} \put(1.6,10.8){$P^a(t)+\delta$}
\put(1.6,9.8){$P^a(t)$} \put(1.6,8.8){$P^a(t)-\delta$}
\put(1.6,7.8){\text{mid price}} \put(17,10.2){$Q^a(t)$}
\put(15,5.2){$Q^b(t)$}
\put(14,10.8){$Q^a_1(t)$}\put(14,11.8){$Q^a_2(t)$}
\put(10,4.8){$Q^b_1(t)$}\put(12,3.8){$Q^b_2(t)$}
\put(7,9.5){\vector(0,1){0.4}}\put(7,6.5){\vector(0,-1){0.4}}
\multiput(7,6.6)(0,0.5){6}{\line(0,1){0.3}}
\put(6.8,6.1){\line(1,0){0.4}}\put(6.8,9.9){\line(1,0){0.4}}
\put(7.5,8.5){\text{Spread }$P^a(t)- P^b(t)$}
%% spread
\multiput(5,8)(0.5,0){42}{\line(1,0){0.3}}
\multiput(16,10)(0.5,0){20}{\line(1,0){0.3}}
\multiput(14,6)(0.5,0){24}{\line(1,0){0.3}}
\put(20,9.5){\vector(0,1){0.4}}\put(20,6.5){\vector(0,-1){0.4}}
\put(20,7.5){\vector(0,1){0.4}}\put(20,8.5){\vector(0,-1){0.4}}
\put(19.8,6.1){\line(1,0){0.4}}\put(19.8,9.9){\line(1,0){0.4}}
\put(19.8,8.1){\line(1,0){0.4}}\put(19.8,7.9){\line(1,0){0.4}}
\multiput(20,6.6)(0,0.5){2}{\line(0,1){0.3}}
\multiput(20,8.6)(0,0.5){2}{\line(0,1){0.3}} \put(21,8.6){\text{buy
}$2\delta$\text{ lower}} \put(21,6.6){\text{sell }$2\delta$\text{
higher}}
%% dark pool
\put(4.9,7){\line(1,0){0.2}} \put(4.9,8){\line(1,0){0.2}}
\put(4.9,9){\line(1,0){0.2}} %% orders
\put(13.7,6.8){\text{arrival}}
 \linethickness{0.04cm}
%% limit buy
%\color{red}
\put(5,3){\line(1,0){4}} \put(5,4){\line(1,0){6}}
\put(5,5){\line(1,0){4}} \put(5,6){\line(1,0){9}}
\put(9,7){\line(1,0){4}} \put(9,7){\vector(-1,0){0.4}}
%% limit sell
\color{darkgray}\put(5,10){\line(1,0){11}} \put(5,11){\line(1,0){8}}
\put(5,12){\line(1,0){8}} \put(5,13){\line(1,0){2}}
\end{picture}
\caption{A limit order book}
\end{figure}
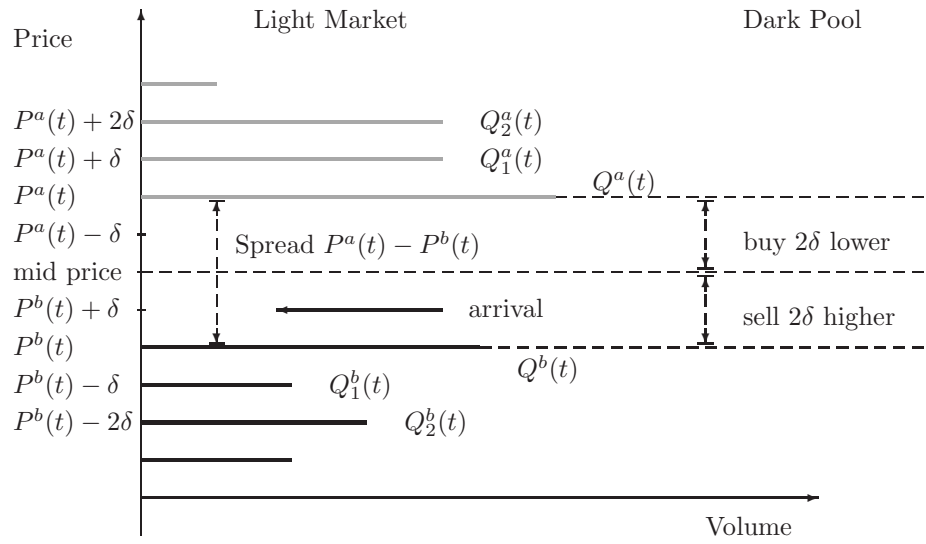
\end{center}

%%%%%%%%%%%%%%%%%%%%%%%%%%%%%%%%%%%%%
\newpage
\begin{center}
\begin{figure}[!h]\label{plot price}
\includegraphics[width=14cm,height=16cm]{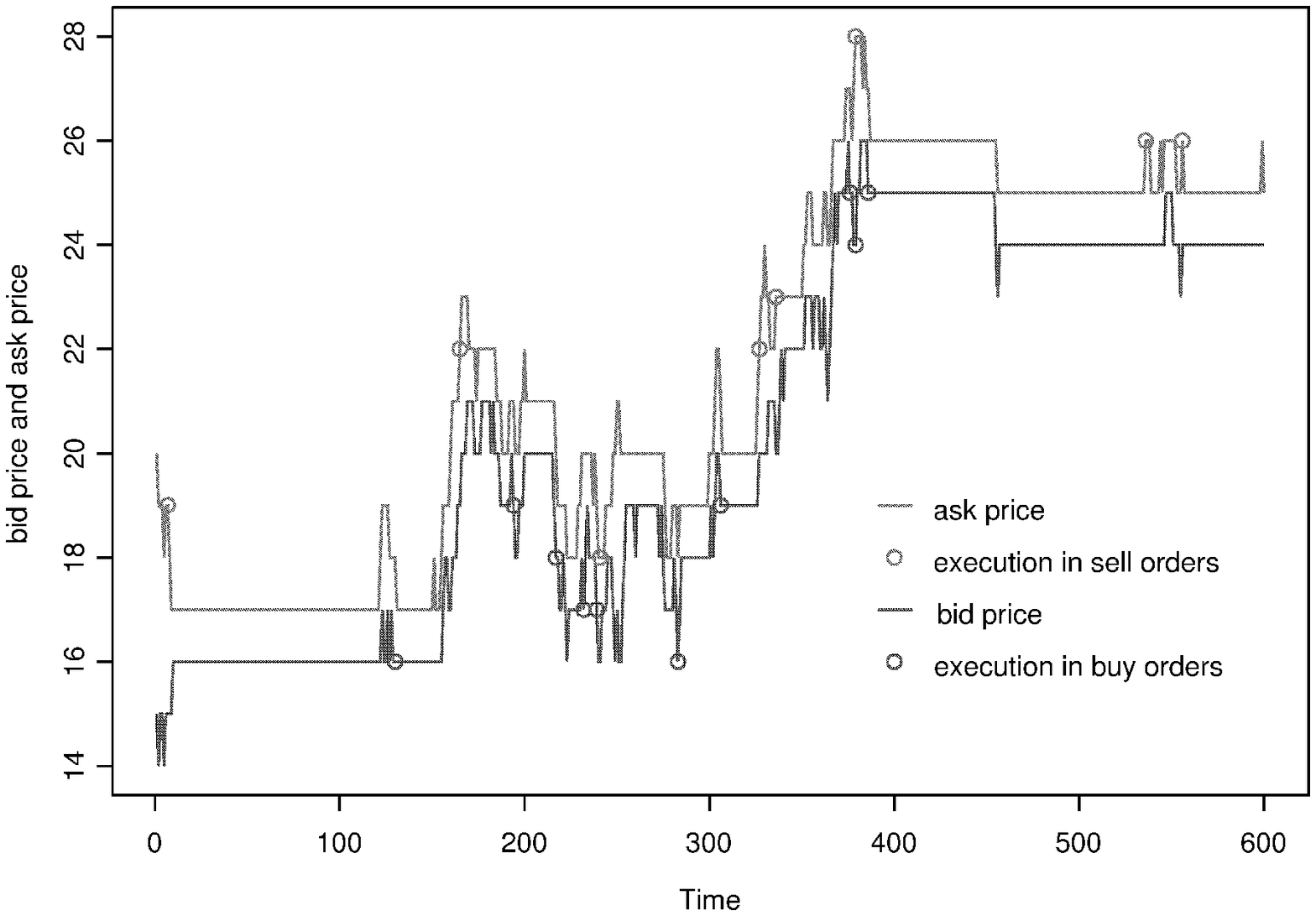}
\caption{Simulation of the bid and ask prices}
\end{figure}
\end{center}
%%%%%%%%%%%%%%%%%%%%%%%%%%%%%%%%%%%%%

%%%%%%%%%%%%%%%%%%%%%%%%%%%%%%%%%%%%%
\newpage
\begin{center}
\begin{figure}[!h]\label{plot volume}
\includegraphics[width=14cm,height=16cm]{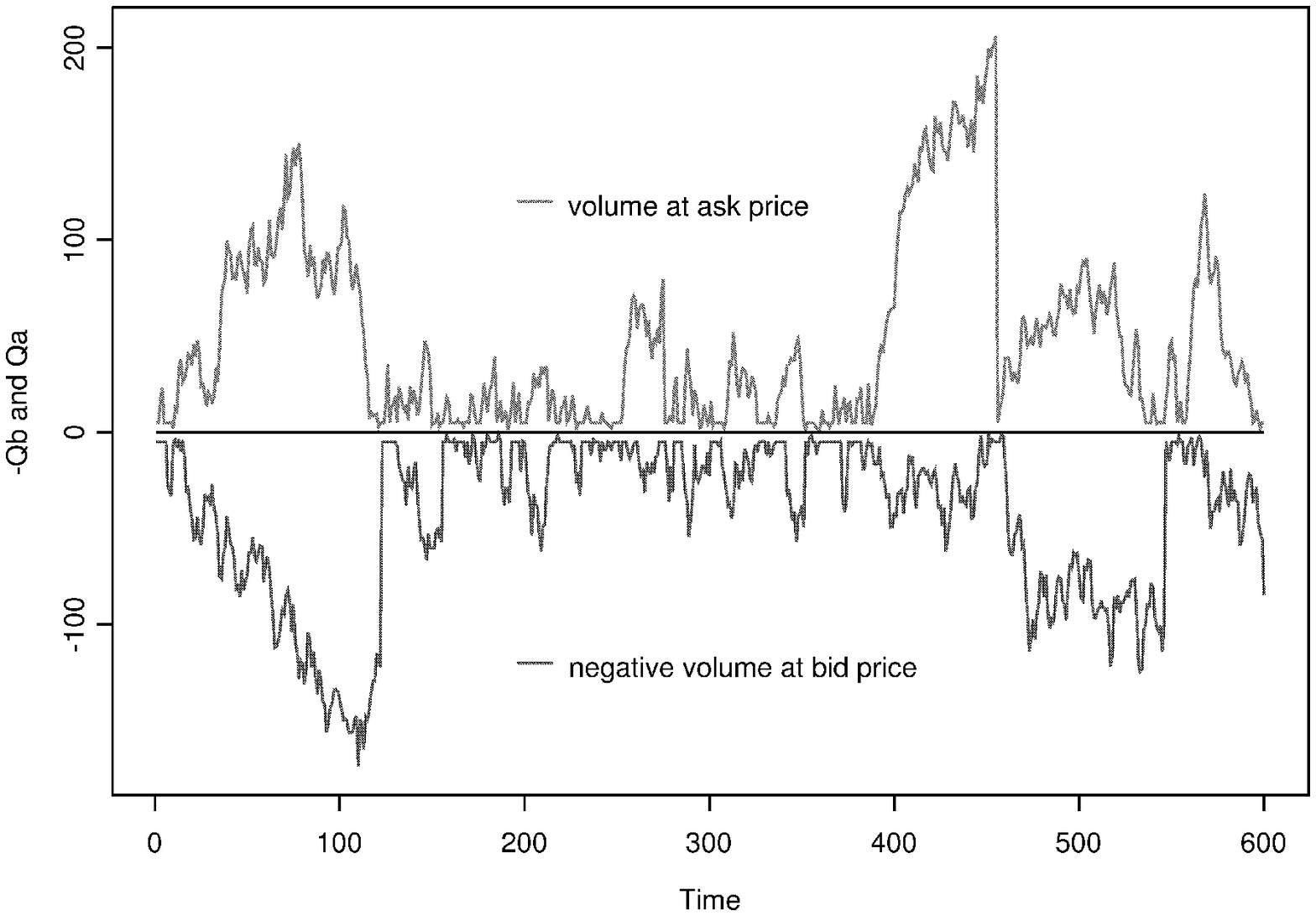}
\caption{Simulation of the volumes}
\end{figure}
\end{center}
%%%%%%%%%%%%%%%%%%%%%%%%%%%%%%%%%%%%%

%%%%%%%%%%%%%%%%%%%%%%%%%%%%%%%%%%%%%
\newpage

\subsection*{Figures from Section \ref{sec sol}}
\setcounter{section}{6} \setcounter{figure}{0}

\begin{center}
\begin{figure}[!h]\label{plot price reg}
\includegraphics[width=14cm,height=10cm]{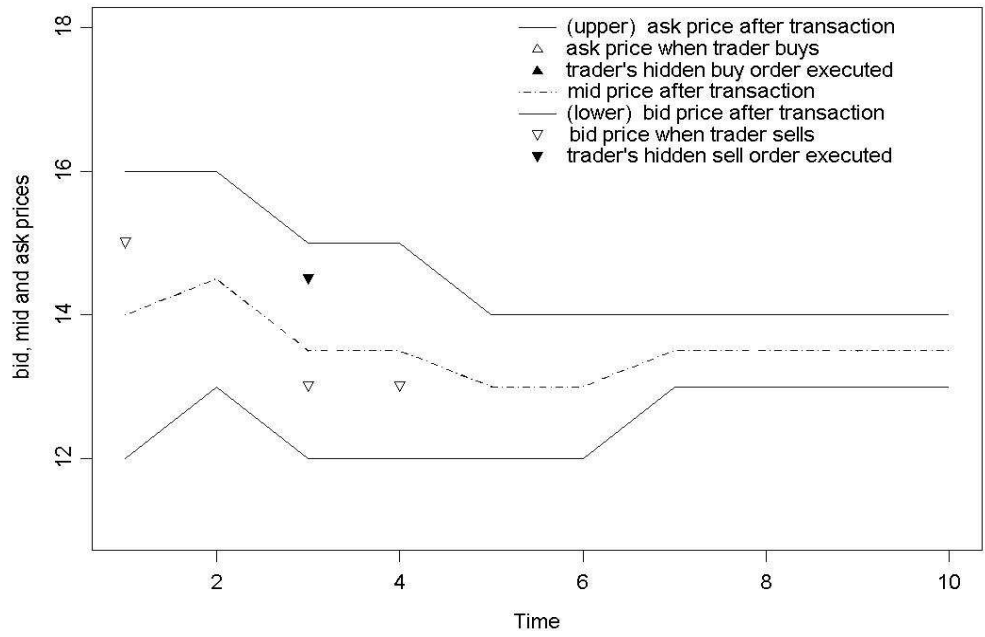}
\caption{Simulated path of prices, regular trader}
\end{figure}
\end{center}
%%%%%%%%%%%%%%%%%%%%%%%%%%%%%%%%%%%%%

%%%%%%%%%%%%%%%%%%%%%%%%%%%%%%%%%%%%%
\newpage

\begin{center}
\begin{figure}[!h]\label{plot price sys1}
\includegraphics[width=14cm,height=10cm]{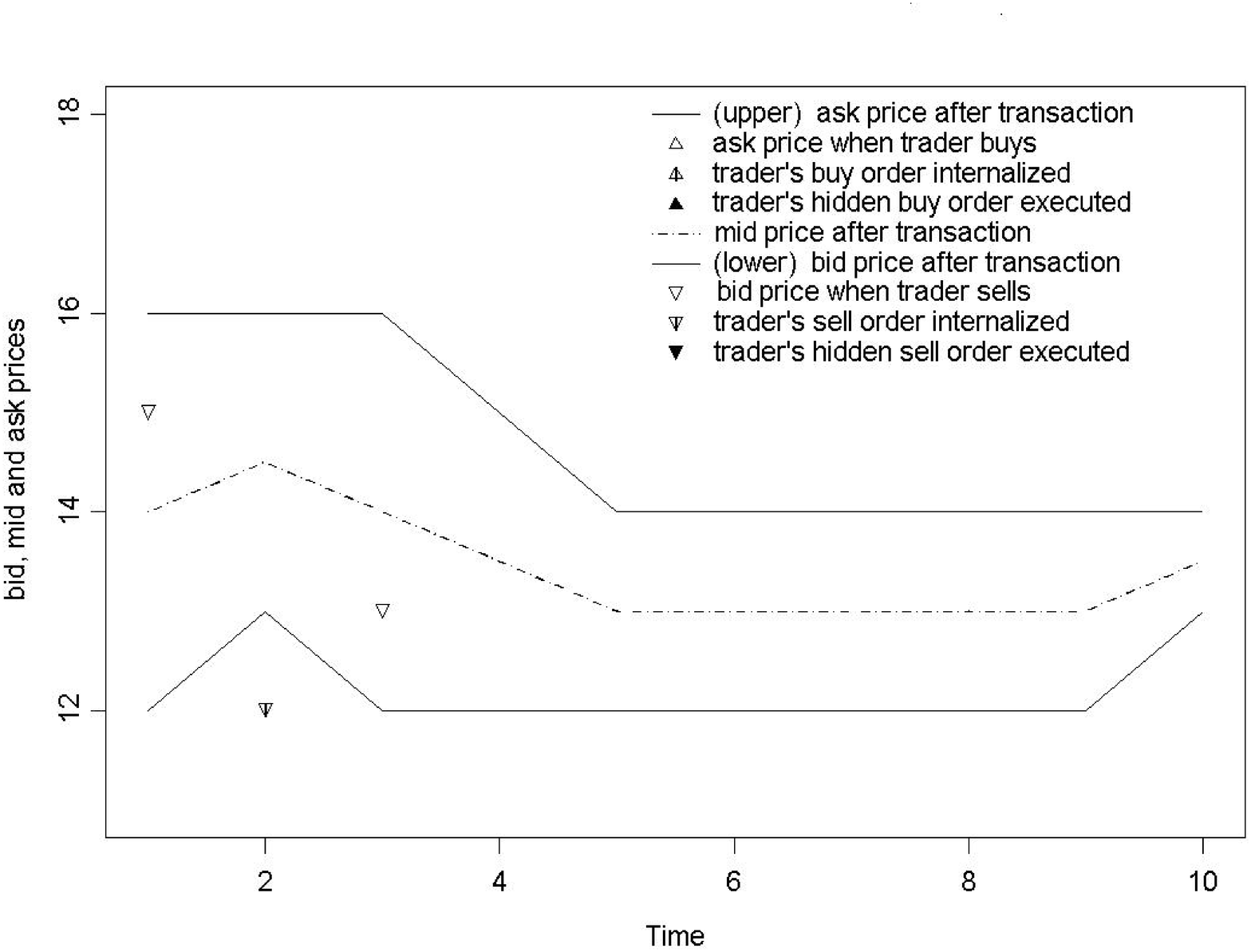}
\caption{Simulated path of prices, systemic internalizer}
\end{figure}
\end{center}
%%%%%%%%%%%%%%%%%%%%%%%%%%%%%%%%%%%%%

%%%%%%%%%%%%%%%%%%%%%%%%%%%%%%%%%%%%%
\newpage

\begin{center}
\begin{figure}[!h]\label{plot price sys2}
\includegraphics[width=14cm,height=10cm]{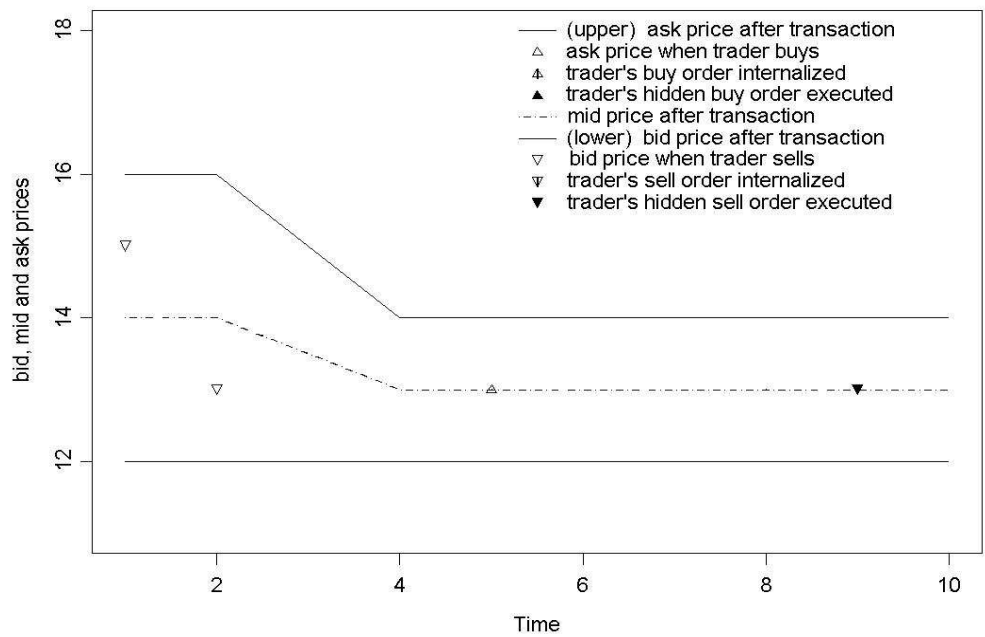}
\caption{Another simulated path of prices, systemic internalizer}
\end{figure}
\end{center}
%%%%%%%%%%%%%%%%%%%%%%%%%%%%%%%%%%%%%

%%%%%%%%%%%%%%%%%%%%%%%%%%%%%%%%%%%%%
\newpage

\begin{center}
\begin{figure}[!h]\label{plot diff}
\includegraphics[width=14cm,height=10cm]{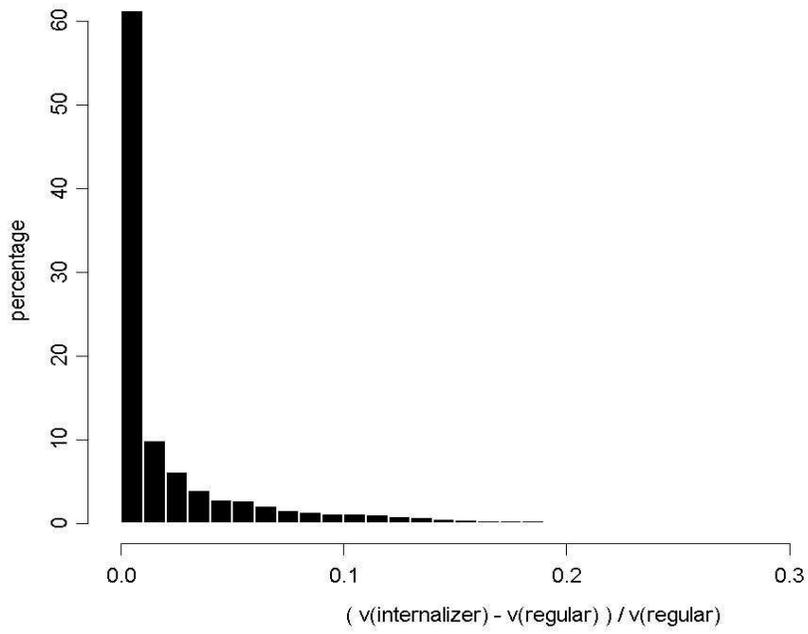}
\caption{Relative difference in best expected profits}
\end{figure}
\end{center}
%%%%%%%%%%%%%%%%%%%%%%%%%%%%%%%%%%%%%

\end{document}